\documentclass[aps,prx,showpacs,twocolumn,superscriptaddress,11,longbibliography]{revtex4-2}

\usepackage{graphicx}
\usepackage{amsmath}
\usepackage{bm}
\usepackage{hyperref}
\hypersetup{%
	pdftitle={Graphene as a source of entangled plasmons},%
	pdfauthor={Zhiyuan Sun, D.N. Basov, and M.M. Fogler},%
	pdfpagemode={UseNone},%
	pdfstartview={FitH},%
	breaklinks=true,%
	citecolor=blue,%
	colorlinks=true,%
	linkcolor=blue,%
	urlcolor=blue
}

\newcommand{\unit}[1]{\,\mathrm{#1}} 
\newcommand{\real}[1]{\mathrm{Re}[#1]} %
\newcommand{\equa}[1]{Eq.~\eqref{#1}} %
\newcommand{\secref}[1]{Sec.~\ref{#1}} %
\newcommand{\fig}[1]{Fig.~\ref{#1}} %

\graphicspath{{./}}

\begin{document}

\title{Graphene as a source of entangled plasmons}

\author{Zhiyuan Sun}
\affiliation{Department of Physics, University of California San Diego, 9500 Gilman Drive, La Jolla, California 92093}
\affiliation{Department of Physics, Columbia University,
	538 West 120th Street, New York, New York 10027}
\affiliation{Department of Physics, Harvard University, 17 Oxford Street, Cambridge, Massachusetts 02138}

\author{D. N. Basov}
\affiliation{Department of Physics, Columbia University,
	538 West 120th Street, New York, New York 10027}

\author{M. M. Fogler}
\affiliation{Department of Physics, University of California San Diego, 9500 Gilman Drive, La Jolla, California 92093}

\date{\today}

\begin{abstract}
We analyze nonlinear optics schemes for generating pairs of quantum entangled plasmons in the terahertz-infrared range in graphene.
We predict that high plasmonic field concentration and strong optical nonlinearity of monolayer graphene
enables pair-generation rates much higher than those of  conventional photonic sources.
The first scheme we study is spontaneous parametric down conversion in a graphene nanoribbon.
In this second-order nonlinear process a plasmon excited by an external pump splits into a pair of plasmons, of half the original frequency each, emitted in opposite directions.
The conversion is activated by applying a dc electric field
that induces a density gradient or a current across the ribbon.
Another scheme is degenerate four-wave mixing where the counter-propagating plasmons are emitted at the pump frequency.
This third-order nonlinear process does not require a symmetry-breaking dc field.
We suggest nano-optical experiments for measuring position-momentum entanglement of the emitted plasmon pairs.
We estimate the critical pump fields at which the plasmon generation rates
exceed their dissipation, leading to parametric instabilities.
\end{abstract}

\maketitle


\section{Introduction}
\label{sec:Introduction}
Graphene plasmonics has emerged as a platform for realizing strong light-matter interaction on ultrasmall length scales~\cite{Koppens.2011,Basov2014}.
Launching, detection, and manipulation of plasmons by near-field probes has been demonstrated at infrared/THz frequencies~\cite{Fei2012, Chen2012, Woessner2015, Shi:2015vq,Ni2016, Lundeberg2017,Ni2018, Wang:2020vs, Dong:2021tn, Zhao:2021vn, Wang:2021ud, Hu:2022vg}.
High concentration of electric field,
a key factor for nonlinear optical phenomena,
has been achieved by exciting plasmons
of wavelength $\lambda_{p}$ as small as few hundred nanometers,
about two orders of magnitude shorter than the vacuum photon wavelength $\lambda_v$.
Plasmonic quality factors up to $Q \sim 130$ 
have been reached~\cite{Ni2018} in encapsulated graphene structures~\cite{Woessner2015,Ni2016},
fulfilling another condition --- low dissipation --- necessary for prominent nonlinear effects.

\begin{figure}
	\includegraphics[width=0.9 \linewidth]{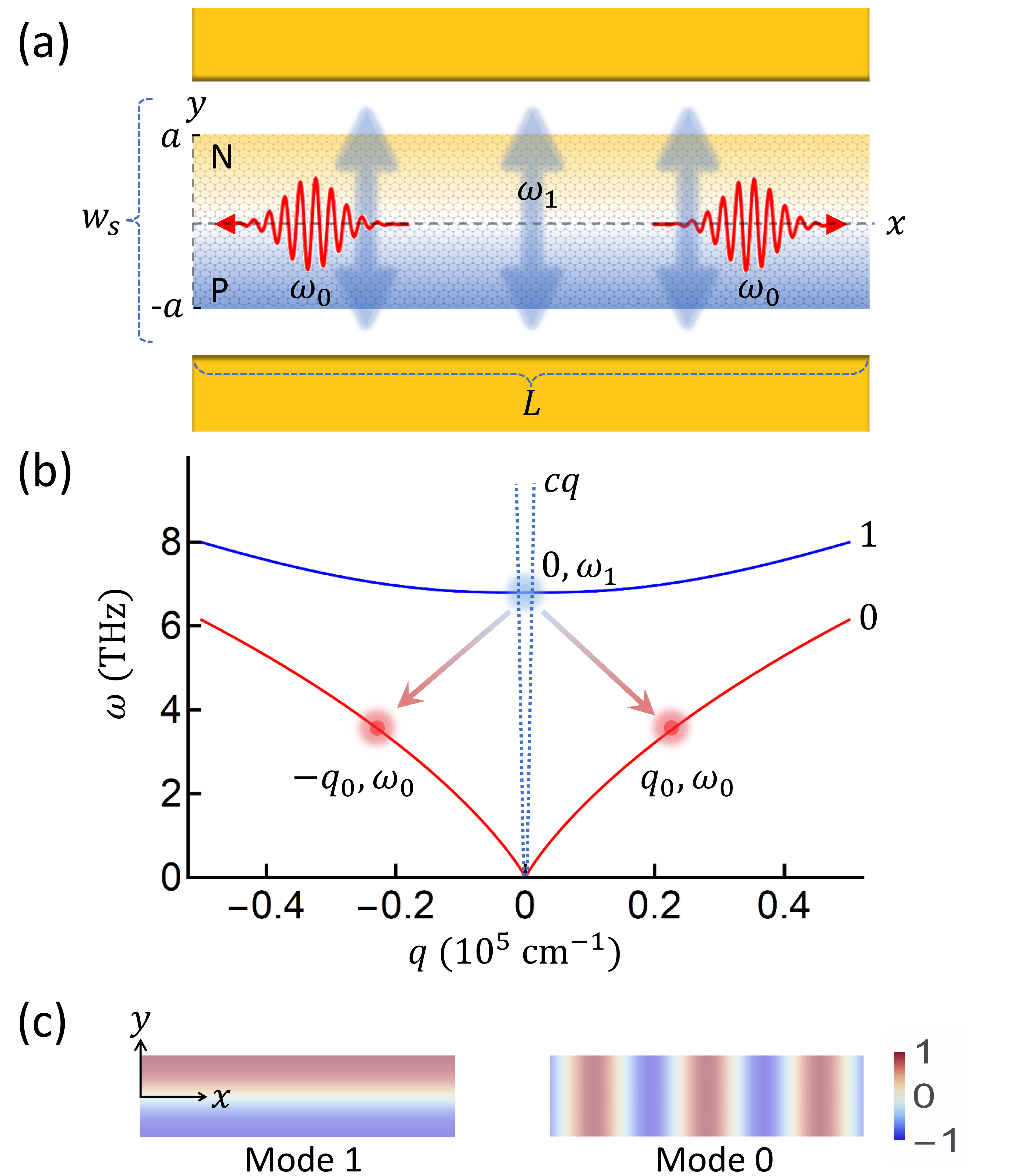} 
	\caption{(Color online)
		(a) Type-A device: a graphene ribbon with a lateral \textit{p}-\textit{n} junction (blue/gold  denoting electron/hole doping) induced by a split gate (golden bars).
		A pump field of frequency $\omega_1$ applied in the $\hat{\mathbf{y}}$-direction (thick arrows)
		generates plasmon pairs of frequency $\omega_0$ propagating in the $\pm\hat{\mathbf{x}}$ direction (red undulating arrows).
		(b) Plasmon dispersion in a ribbon of half-width $a = 200 \unit{nm}$ with the \textit{p}-\textit{n} junction on the midline and the carrier density $n = \pm 10^{12} \unit{cm^{-2}}$ at the edges. The dotted lines form the `light cone' for the dispersion of vacuum photons. The blue dot represents the mode-$1$ plasmon pumped by the incident photon, which then decays into two mode-$0$ plasmons (the two red dots) as shown by the arrows.
		(c) The spatial distribution of the amplitude of the quasi-static electric potential of the two modes.
	}
	\label{fig:pn}
\end{figure}

Numerous theoretical~\cite{Mikhailov2007, Mikhailov2011, Manzoni, Glazov2014, Cheng2014, Wang2016, Mikhailov2017, Mikhailov2017a, Sun2018a, Sun2018b} and a few experimental~\cite{Constant2016, Yao2018} works have explored manifestations of nonlinear coupling of light to graphene in the context of conventional far-field optics.
However, both nonlinear~\cite{Kroo2008,Gullans2013,Kauranen:2012tb, Cox:2017uq, Sun2018a} and quantum~\cite{Tame2013, Gullans2013,Reserbat-Plantey.2021} effects could be more pronounced in the near-field domain
because of plasmonic field concentration.
Plasmon interaction phenomena we study in this paper include the  parametric down conversion (PDC) and  four-wave mixing (FWM)~\cite{Boyd.2008}.
Spontaneous PDC and FWM can generate entangled pairs of plasmons,
similar to how the usual photonic spontanenous PDC produces entangled pairs of photons~\cite{Reid2009,Pan.2012,Orieux.2017,Kwiat1995,Wang.2019,DeRossi.2002,Steiner2021uep,Javid.2021}.

We propose that a graphene ribbon containing an electrostatically induced \textit{p}-\textit{n} junction~\cite{Fogler2008, Vicarelli:2012vg, Woessner2015, Woessner2017, Alonso-Gonzalez:2017vt, Castilla:2019tf}, see Fig.~\ref{fig:pn}(a),
can be a highly efficient PDC source.
The plasmon spectrum of such a system, which we will call the device of type A,
consists of multiple continuously dispersing subbands. 
The lowest subband (labeled ``$0$'') is gapless and the next one (labeled ``$1$'') is gapped due to transverse confinement in the ribbon, see Fig.~\ref{fig:pn}(b).
In the simplest picture,
the PDC process is
splitting of a mode-$1$ plasmon of frequency $\omega_1$ and momentum $q_1 = 0$
into two mode-$0$ plasmons of frequency $\omega_0 = \omega_1 / 2$ and momenta $\pm q_0$ that
propagate away in opposite directions along the ribbon.
The energy-momentum conservation (phase matching) in this conversion is guaranteed without any fine-tuning of the device.
This advantage of the counterpropagating PDC scheme has been
previously exploited in photonic waveguides~\cite{Ding.1995, DeRossi.2002, Orieux.2017, Saravi.2017}.
In a more precise description, frequencies $\omega_0$ and so the momenta $q$
of the outgoing plasmons have some uncertainty.
The spontanenous PDC generates a superposition of states of different momentum:
\begin{equation}
|\psi\rangle=	\int dq f(q)\, |{-q} \rangle\, |{+q} \rangle
	\label{eqn:SPDC}
\end{equation}
where $|{-q} \rangle\, |{+q} \rangle$ means a two plasmon state with one plasmon at momentum $-q$ and the other at $+q$, see Sec.~\ref{sec:entanglement_measure}.
This superposition is not factorizable, which implies that
the pairs are position-momentum entangled, similar to particles
in the original Einstein, Podolsky, and Rosen (EPR) paper~\cite{Reid2009}.
The amplitude function $f(q)$ in Eq.~\eqref{eqn:SPDC} is peaked at $q_0$ and has a characteristic width $\delta q \sim \gamma / v_0$ where
\begin{equation}
 \gamma = \omega_0 / Q
 \label{eqn:gamma}
\end{equation}
is the plasmon damping rate~\cite{Ni2018} and $v_0 = d\omega_0 / d q_0$ is the mode-$0$ plasmon group velocity.

The role of the split-gate represented by the golden bars in Fig.~\ref{fig:pn}(a) is two-fold.
First, it induces a gradient of carrier concentration in graphene
when a dc voltage is applied between the two parts of the gate.
The \textit{p}-\textit{n} junction is created when this voltage is high enough~\cite{Woessner2017}.
Such a system lacks inversion symmetry and therefore the PDC is allowed.
Second, the split-gate serves as an optical antenna amplifying the incident pump field~\cite{Woessner2017,Jiang:18}
that excites mode-$1$ plasmons.
The coupling is facilitated by a large transverse dipole moment
of the mode-$1$ plasmons and a strong field enhancement $\lambda_v / \lambda_p \gg 1$ of the antenna.

Note that the inversion symmetry can also be broken by a dc current
across a uniformly doped ribbon,
which we refer to as the device of type B.
%
%
In contrast, the FWM is a third-order nonlinear effect
which does not require symmetry breaking fields. The degenerate FWM, in which the two pump photons of frequency $\omega$ are converted into a pair entangled plasmons of the same frequency, can take place in either a nanoribbon or a more typical larger-area graphene sheet, which are devices of type C.

Plasmonic effects similar to those we explore here have been considered
in a few recent theoretical papers. For example, plasmonic sum-frequency generation in graphene nanoflakes~\cite{Manzoni} is based on the nonlinear mechanism similar to that underlying our plasmonic PDC.
However, since the plasmon spectrum of a flake is discrete, 
only flakes of certain shapes can fulfill the energy-momentum conservation constraint for the sum (or difference) frequency generation. 
As mentioned above, in a long ribbon such kinematic constraints can be 
met without delicate fine-tuning.
Another related work studied photon-plasmon difference frequency generation \cite{Yao2018,wolff2020stimulated} and entanglement~\cite{Tokman2016},
which is a phenomenon intermediate between
the usual all-photon PDC~\cite{Reid2009} and our fully plasmonic PDC.

The remainder of the paper is organized as follows. In Sec.~\ref{sec:main_results}, we summarize our main results for the type-A device.
In the last part of that section we discuss possible  experiments that can probe
the plasmon entanglement.
In Sec.~\ref{sec:third_order} we consider the third-order nonlinear effects. The FWM
relevant for the operation of type-C devices is analyzed in Sec.~\ref{sec:four_wave_mixing} and the current induced PDC in type-B devices is examined in Sec.~\ref{sec:spdc_current}.
Section~\ref{sec:Discussion} contains discussion and outlook.
Finally, the summary of the notations and some details of the derivations
are given in the Appendix.

\section{Plasmonic PDC in a graphene ribbon}
\label{sec:main_results}

To simplify the analysis of a type-A device, we assume that
the edges of the ribbon $y = \pm a$ are not too close to those of the split-gate,
so that the dc electric field created by the gate is approximately constant across the ribbon.
The induced density response of graphene can be  
modeled~\cite{Fogler2008} assuming the local chemical potential is linear in $y$:
\begin{equation}
\mu(y) = \frac{a + y}{2a} \mu_t + \frac{a - y}{2a} \mu_b\,.
\label{eqn:mu}
\end{equation}
If the chemical potentials $\mu_t$ and $\mu_b$ at the two edges of the ribbon are opposite in sign, a \textit{p}-\textit{n} junction forms.
If $\mu_t = -\mu_b$, the junction is located on the midline $y = 0$ of the ribbon.

The plasmon frequency dispersion computed numerically as a function of momentum $q$ along the strip is shown in Fig.~\ref{fig:pn}(b) (For related analytical results, see Ref.~\onlinecite{Mishchenko2010}).
The lowest-frequency branch, which we refer to as mode $0$, is gapless:
\begin{align}
\omega(q) = \xi_v \sqrt{\alpha_g a k_F} v_F q = \xi_v v_a q \,.
\label{eqn:acoustic_dispersion}
\end{align}
Here
\begin{equation}
	v_a = \sqrt{(e^2 /\hbar) v_F k_F a} \sim v_F \sqrt{k_F a}\,,
	\label{eqn:v_a}
\end{equation}
is a characteristic scale of the plasmon velocity,
$v_F \approx c / 300$ is the Fermi velocity,
$\alpha_g = e^2 / (\hbar v_F)$ is the dimensionless strength of the Coulomb interaction, and $k_F$ denotes the Fermi momentum corresponding to the maximum
absolute chemical potential $\max(|\mu_t|, |\mu_b|)$ in the ribbon.
Dimensionless parameter $\xi_v \sim 1$ 
is a slow (logarithmic) function of $q$. It also depends on the doping profile of the ribbon.
The dispersion law~\eqref{eqn:acoustic_dispersion} is similar to that of one dimensional (1D) plasmons~\cite{Shi:2015vq, Wang:2020vs, Wang:2021ud}.

The first gapped mode, mode $1$, can be viewed as a linear combination of 2D plasmons of momenta $q \hat{\mathbf{x}} \pm q_y \hat{\mathbf{y}}$ with $q_y \sim 1 / a$.
The frequency of this mode at $q = 0$ is
\begin{align}
\omega_1 = \xi_1 \sqrt{2\alpha_g v_F^2 k_F/a} \,,
\label{eqn:gapped_dispersion}
\end{align}
where $\xi_1 \sim 1$ is another dimensionless shape factor.
The electric potential profile of the $\omega_1$ mode [Fig.~\ref{fig:pn}(c)] indicates that mode-$1$ excitations have a nonzero dipole moment in the $y$-direction.
Thus, they can be resonantly excited by a $y$-polarized pump field~\cite{Hu:2017wi}.

In this paper we describe the mode-$1$ to mode-$0$ PDC in the quantum language.
However, there is a complementary classical picture, which is as follows.
Plasmonic oscillations of mode-$1$ modulate the total carrier density of the ribbon viewed as a one dimensional conductor.
This in turn modulates the frequencies of mode-$0$ plasmons and causes their parametric excitation.
In the standard theory of optical parametric amplification~\cite{Boyd.2008} (see also Appendix~\ref{sec:OPA}), one of the generated mode-$0$ plasmon is referred to as the idler and the other  as the signal.
The idler is assumed to have a finite spectral power, e.g., due to equilibrium fluctuations in the beginning of the process. The signal is then produced from the pump and the idler by the difference frequency generation (DFG).

For weak pump fields $E$, the pair generation rate grows linearly with the pump power, see Fig.~\ref{fig:generation_growth_rate}.
As $E$ becomes stronger, a parametric instability is reached at some critical $E$.
At even stronger pump field,
the signal exhibits an exponential growth in the undepleted pump approximation.
In Sec.~\ref{sec:ISADR} below we evaluate the dependence of the pair generation rate and the critical pump field on experimental parameters, such as the ribbon width and its doping profile.

\subsection{PDC Hamiltonian}
\label{sec:ISADR}

\begin{figure}
\includegraphics[width=0.9 \linewidth]{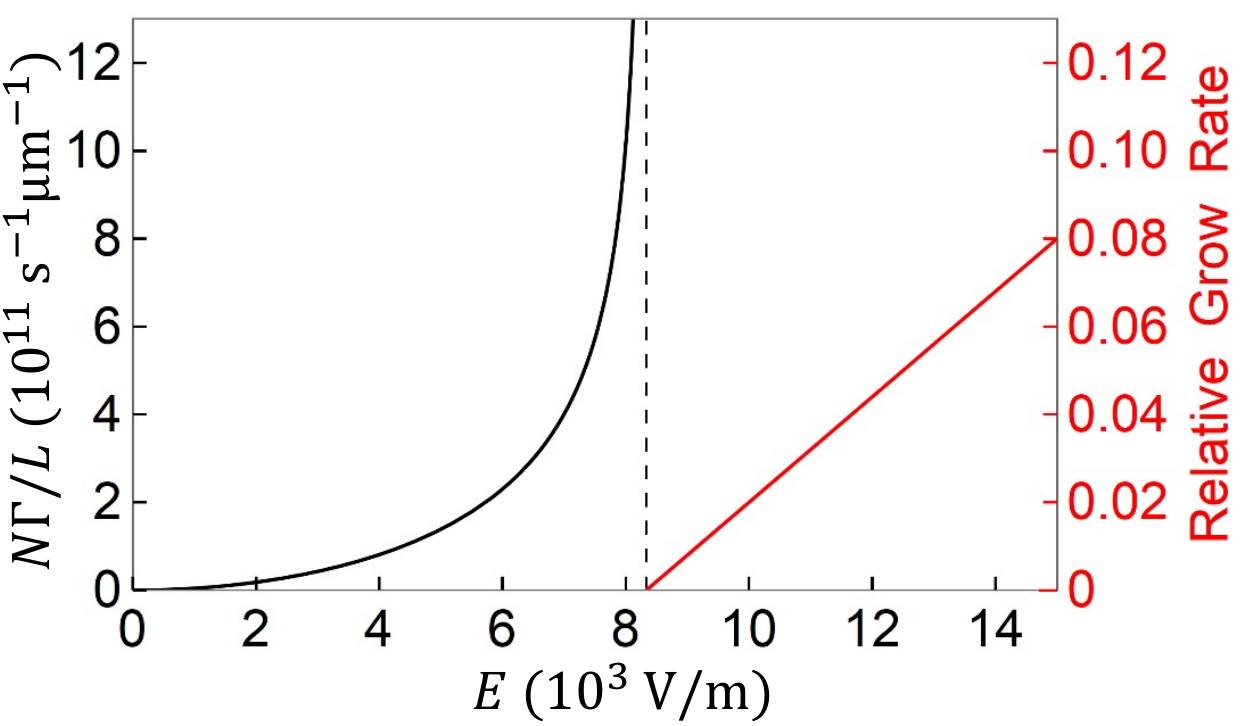} 
\caption{Transition from stable to unstable regime of plasmon-pair generation in the type-A device as pump field is increased. The black curve is the rate of generation of plasmon pairs per unit length. The red curve is the relative growth rate $(\kappa - \gamma) / \omega_0$ in the unstable regime. The plasmon damping rate is assumed to be $\gamma = 0.1 \omega_0$. Other parameters are from \fig{fig:pn} and the temperature is $T=300 \unit{K}$. 
}
\label{fig:generation_growth_rate}
\end{figure}

The second-order nonlinear conductivity \cite{Mikhailov2011, Tokman2016, Cheng2016, Sun2018a}  leads to interaction between the plasmons of modes $0$ and $1$.
This interaction can be modeled by the Hamiltonian $H = H_0 + H^{(2)}$:
\begin{align}
H_0 &= \hbar \omega_1 (a^\dagger_1 a^{\phantom\dagger}_1 + 1/2)
 + \sum_q\hbar \omega(q) (a_{q}^\dagger a_q + 1/2) \,, \notag \\
H^{(2)} &=  
\sum_q g^{(2)}_q a_1 a_q^\dagger a_{-q}^\dagger + \text{c.c.}
\label{eqn:hamiltonian}
\end{align}
where $a_1$, $a_q$ are the annihilation operators for mode-$1$ at zero momentum and mode-$0$ of momentum $q$, respectively.
The interaction Hamiltonian $H^{(2)}$ leads to pair generation in the weak coupling regime ($g^{(2)} a_1 Q/\omega \ll 1$) and two-mode squeezing in the strong coupling regime ($g^{(2)} a_1 Q/\omega > 1$), as shown in Fig.~\ref{fig:generation_growth_rate}. Since we study the PDC on resonance, $\omega(q_0)= \omega_0 =\omega_1/2$, we can treat the interaction strength $g^{(2)}$ as a constant, $g^{(2)} = g^{(2)}_{q_0}$.
As shown in Appendix~\ref{sec:derivation},
\begin{align}
g^{(2)} = \xi_g \frac{\pi^{1/2}}{2^{13/4}} \frac{ e^{1/2}  \hbar^{3/4} v_F^{3/4} }{k_F^{5/4} a^{7/4} L^{1 / 2}} \,,
\label{eqn:g}
\end{align}
where $L$ is the length of the gated part of the ribbon (Fig.~\ref{fig:pn})
and dimensionless factor $0 \leq \xi_g \leq 1$ depends on the carrier density profile across the ribbon.

The Hamiltonian $H^{(2)}$ governs the spontantous decay of the mode-$1$ plasmon into a pair of mode-$0$ plasmons with momenta $q$ and $-q$.
In the weak-coupling regime,
the decay rate $\Gamma$ can be calculated using Fermi's golden rule (see Appendix~\ref{sec:decay_rate})
\begin{align}
\Gamma = \frac{|g^{(2)}|^2 }{\hbar^2} \frac{L}{v_0} \left(n_0 + \frac12 \right)  ,
\label{eqn:decay_rate}
\end{align}
where $n_0$ is the occupation number of mode-$0$ plasmons of frequency $\omega \simeq \omega_0$; in thermal equilibrium, $n_0 = 1/(e^{\hbar \omega_0 / T} - 1)$ at temperature $T$.
Normalizing $\Gamma$ to the plasmon frequency $\omega_0$,
we obtain the dimensionless decay rate of mode-$1$ plasmons:
\begin{align}
\frac{\Gamma}{\omega_1} =
\xi_{\text{df}} \frac{\Gamma_0}{\omega_1}  
,\quad
\frac{\Gamma_0}{\omega_1}  =  \frac{\pi}{2^8} \frac{1}{\sqrt{\alpha_g}} \frac{1}{(k_F a)^{7/2} }  \,,
\label{eqn:decay_factor}
\end{align}
where $\xi_{\text{df}} \sim 1$ is another dimensionless factor.
The dependence of $\xi_{\text{df}}$ on the carrier density profile across the ribbon is plotted in Fig.~\ref{fig:shape_factor}.
For a ribbon with the carrier density $n = \pm 10^{12} \unit{cm^{-2}}$ at the edges and a half-width of $a = 200 \unit{nm}$, we find $\omega_1 / (2\pi) \approx 5 \unit{THz}$ and
$\Gamma_0 / \omega_1 \sim 10^{-8}$.
Therefore, due to PDC alone
a mode-$1$ plasmon decays into a mode-$0$ plasmon pair every $\Gamma_0^{-1} \sim 10^{-5} \unit{s}$.
If the plasmon $Q$-factor due to other damping channels (phonon and impurity scattering~\cite{Ni2018}) is $Q = \omega_0 / \gamma \sim 10$, the PDC efficiency is
$\Gamma_0 / \gamma \sim 10^{-7}$,
which is much higher than in available photonic PDC devices \cite{DeRossi.2002, Orieux.2011}.
Furthermore, this efficiency scales as $a^{-4}$, and so can be higher still in a narrower strip.

\subsection{Pumping mode-$1$}
\label{sec:Pumping}

\begin{figure}[t]
	\includegraphics[width=0.9 \linewidth]{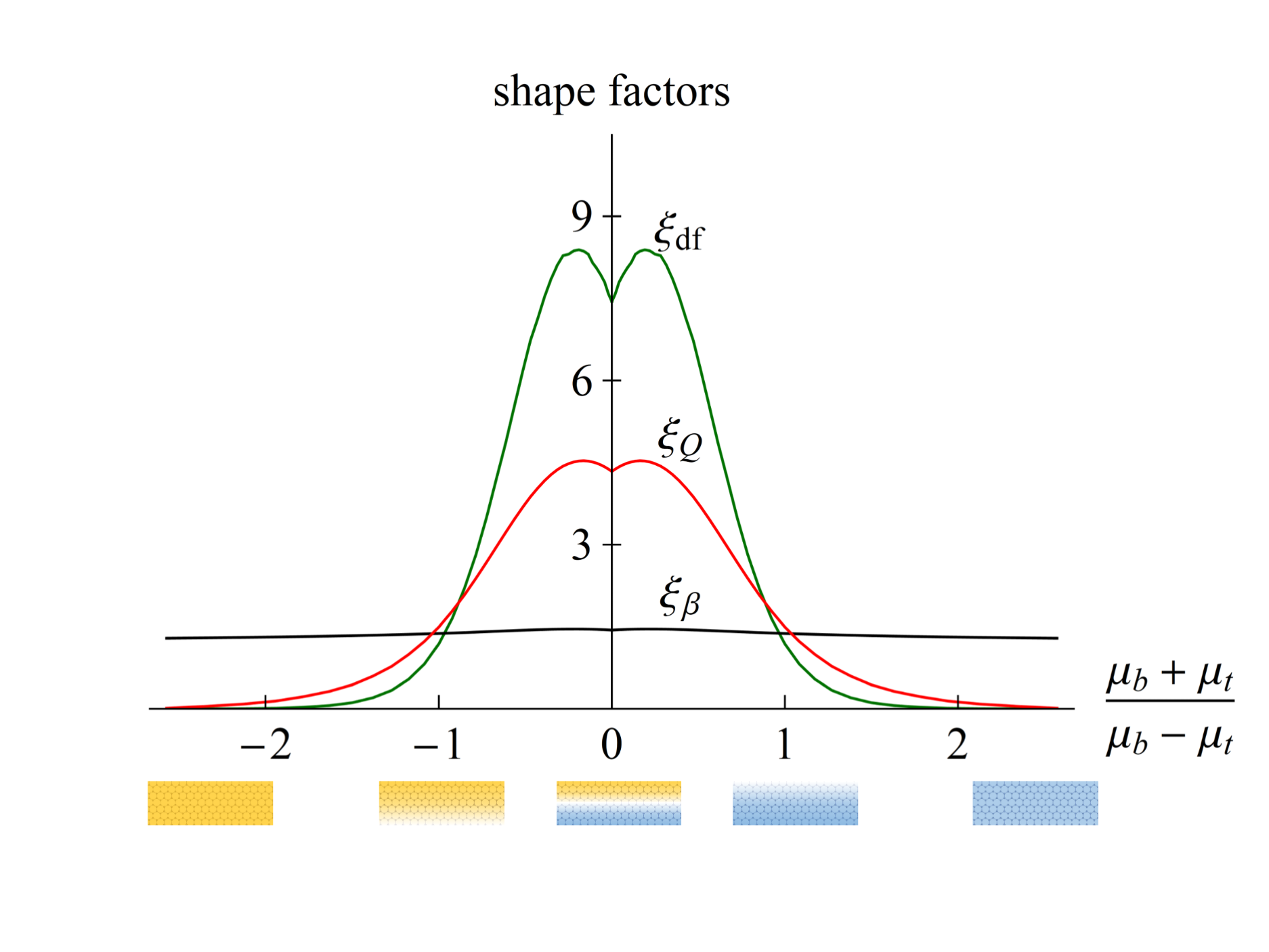} 
	\caption{The shape factors for type-A device as functions of the doping profile parameter $s_d = (\mu_b + \mu_t)/(\mu_b-\mu_t)$ at temperature $T=300\unit{K}$. The green, black and red curves are the decay rate shape factor $\xi_{\text{df}}$, the occupation number shape factor $\xi_\beta$ and the shape factor $\xi_Q$ for the relative subharmonic growth rate (defined in Appendix~\ref{appendix:notations}).
Each colored stripe below the horizontal axis illustrates the corresponding graphene ribbon with blue/gold representing electron/hole doping.}
	
	\label{fig:shape_factor}
\end{figure}

In a uniform external ac electric field $E$ in the $y$-direction, the gapped mode-$1$, as a harmonic oscillator, is driven to a coherent state $|\beta\rangle=e^{\beta a_1^\dagger - (\beta^2 / 2)} |0\rangle$. 
Specifically, the magnitude of the charge density profile of the mode is related to the external pump field as
\begin{align}
	\rho = \frac{ \frac{\sigma(\omega)}{\omega} \frac{\omega_1}{\sigma(\omega_1)}}{(1 -  \frac{\sigma(\omega)}{\omega} \frac{\omega_1}{\sigma(\omega_1)}) } \frac{\alpha}{S} E \,
	\label{eqn:EOM_source}
\end{align}
where $\sigma(\omega)$ is the two dimensional optical conductivity and $\alpha/S$ is an $O(1)$ constant determined by the doping profile of the ribbon, see Appendix~\ref{appendix:nano_structure}.
On resonance, from its relation to $\rho$ (see Appendices~\ref{sec:Quantization} and \ref{appendix:ribbon}), $\beta$ is found  to be
\begin{align}
\beta &= \xi_\beta \beta_0
,\quad
\beta_0 = \frac{2^{\frac34}}{\pi^{\frac12}}\, Q\, \sqrt{ \frac{L}{a}} \frac{e a E}{\left(\frac{e^2}{a} \right)^{\frac{3}{4}} \varepsilon_F^{\frac{1}{4}}} 
\label{eqn:beta}
\end{align}
where $\xi_\beta \sim 1$ is a shape factor shown in Fig.~\ref{fig:shape_factor}. The occupation of the gapped mode modifies the pair-generation rate to 
\begin{align}
	R = N \Gamma 	\propto  E^2 a^{-3/2} L,
	\quad
	N = \beta^2
	\propto E^2 a^{5/2} L  \,.
	\label{eqn:R}
\end{align}
For $n = 10^{12} \unit{cm^{-2}}$, $a = 200 \unit{nm}$, $L = 1 \unit{\mu m}$, $E = 10^3 \unit{V/cm}$ and a  plasmon quality factor of  $Q \sim 10$ for ultra clean samples at room temperature \cite{Yan.2013, Woessner2015, Ni2016, Ni2018}, we have $\beta_0 = 15$, indicating an average occupation number of $N \sim 400$, thus boosting the generation rate to $R \sim 4 \times 10^9 \unit{s^{-1}}$. It means that in a nanosecond-long pulse, there are four pairs of subharmonic plasmons generated  while there are $\sim 10^6$ photons incident onto the ribbon area. This rate is higher than conventional photonic~\cite{DeRossi.2002} and quantum dot~\cite{Wang.2019}  devices. 

As the pump field increases, Fermi's golden rule ceases to be valid.
The pair generation rate can be derived from the classical parametric oscillator theory (Appendix~\ref{sec:OPA})
\begin{align}
R = 
 \frac{N\Gamma}{\sqrt{1-(\kappa /\gamma)}} \,,
 \quad
 \kappa = \frac{\beta g^{(2)}}{\hbar}\,.
\label{eqn:R}
\end{align}
The growth rate $\kappa$ is to be discussed in the next section. The pair generation rate diverges at the instability threshold $\kappa = \gamma$, as shown in Fig.~\ref{fig:generation_growth_rate}.

\subsection{Two mode squeezing and parametric amplification}
\label{sec:Two_mode_squeezing}

The coupling Hamiltonian Eq.~\eqref{eqn:hamiltonian} in the interaction picture reads
$
H_I =  \sum_q g^{(2)}_{q} e^{-i\delta t} a_1 a_q^\dagger a_{-q}^\dagger + c.c.   \,
\label{eqn:hci}
$
where $\delta= \omega_1 - 2\omega_q $ is the frequency mismatch. We focus on the pair of acoustic modes $\omega(q_1)= \omega_0=\omega_1/2$ for which $H_I$ is on resonance:
\begin{align}
H_I =  g^{(2)} a_1 a_q^\dagger a_{-q}^\dagger + c.c.   \,.
\label{eqn:hci1}
\end{align}
When mode-$1$ is pumped into a coherent state, one can replace $a_1$ by its classical value $\beta$. This Hamiltonian generates a two-mode-squeezing time evolution operator \cite{Caves1985} 
\begin{align}
S(\beta g^{(2)} t/\hbar) = e^{-iH_I t} = e^{-i \left( g^{(2)} \beta a_q^\dagger a_{-q}^\dagger + c.c. \right)t/\hbar}   \,,
\label{eqn:time_evolution}
\end{align}
which squeezes the pair of mode-$0$ plasmons at momenta $(q,-q)$.
Therefore, the plasmons are generated as two-mode squeezed states~\cite{Reid2009}. 
The squeezing operator leads to exponential growth of the amplitudes of the observables (e.g., the electric field of the mode) with the growth rate $\kappa=g^{(2)} \beta/\hbar$.
We define the dimensionless relative growth rate 
\begin{align}
\frac{\kappa}{\omega_0} 
= \frac14 \xi_Q Q \frac{\zeta}{\alpha_g} \,,
\quad
 \zeta = \frac{e E}{\varepsilon_F k_F}
\label{eqn:Qg}
\end{align}
that is made only of intensive quantities. Here $\zeta$ is the dimensionless small parameter that controls the second order nonlinear effects of graphene plasmons.  
Note that the relative growth rate depends only on the maximum doping level, the pump electric field, and the dimensionless shape factor $\xi_Q$. Parameter $\xi_Q \sim 1$, plotted in Fig.~\ref{fig:shape_factor}, depends only on the scale invariant doping profile of the ribbon.
The pump induced growth rate $\kappa$ reduces the plasmon damping rate from $\gamma$ to $\gamma - \kappa$, thereby further enhancing the pair generation rate.
The system reaches parametric instability at $\kappa = \gamma$, see Fig.~\ref{fig:generation_growth_rate}. Beyond this threshold, the plasmons are unstable with exponential growth rate $\kappa-\gamma$. When amplifying a classical source as the seed, this effect is known as parametric amplification~\cite{Boyd.2008}.
For $n = 10^{12} \unit{cm^{-2}}$ and $E = 10^4 \unit{V/cm}$, we have $\kappa = 0.12 \omega_0$ in the $p$-$n$ junction regime, high enough to compete with the damping rate $\gamma$. 

The split-gate structure, as an antenna with width in $y$ larger than  the vacuum wave length $\lambda_v$ of the incident light, could enhance the pump field $E_v$ to the total field $E$ by a factor of roughly $F=E/E_v= \lambda_v/(2\pi w_s)$~\cite{Woessner2017,Jiang:18} where $w_s$ is the distance separating the two halves of the gate, see Fig.~\ref{fig:pn}.
For the typical parameters in Fig.~\ref{fig:pn}, one has $\lambda_v = 43 \unit{\mu m}$ at $\omega_1=7 \unit{THz}$ and $w_s$ could be chosen as $0.5 \unit{\mu m}$, rendering the field enhancement factor $F \approx 14$.	 (To increase the working frequency to, e.g., $\omega_1=30 \unit{THz}$ corresponding to $\lambda_v = 10 \unit{\mu m}$, the ribbon width needs to be shrinked to $2a \sim 20 \unit{nm}$.) Therefore, to achieve the plasmon instability regime in Fig.~\ref{fig:generation_growth_rate}, the actual incident field just needs to be of the order of $E_v=10^2$--$10^3 \unit{V/cm}$.
We note that the antenna also amplifies the radiative damping rate $\gamma_R$ of the dipolar active modes (e.g., mode-$1$) of the ribbon by the same factor $F$. For ribbons of size much smaller than the vacuum wavelength, $a, L \ll \lambda_v$, the normalized radiative damping rate $Q_R^{-1}=\gamma_R/\omega_0$ is amplified by the antenna from $Q_R^{-1} \sim a^2 L/\lambda_v^3$ to $a L/\lambda_v^2$, which is still much smaller than $\gamma/\omega_0 \sim 0.1$ \cite{Yan.2013, Woessner2015, Ni2016, Ni2018}
for the parameters given above.
Therefore, the intrinsic damping $\gamma$ (e.g., due to phonons and the electronic system itself) should be the major plasmon dissipation pathway in the system and the effect of the radiative damping can be neglected.

\subsection{Measuring the entanglement}
\label{sec:entanglement_measure}
In general, the plasmon pair state can be expressed as
\begin{equation}
	|\psi \rangle =  \int dp_1 dp_2 f(p_1,p_2) |p_1\rangle\, |p_2\rangle\,
	\,.
	\label{eqn:EPR}
\end{equation}
To quantify the standard deviations, we approximate the wave function by the Gaussian form
\begin{equation}
	f(p_1,p_2) = Z e^{-\frac{(p_1+p_2)^2}{4 \Delta_+^2}-\frac{(p_1-p_2-2p_0)^2}{4 \Delta_-^2}}
	\label{eqn:gaussian}
\end{equation}
where $Z$ is a normalization factor. The EPR state in \equa{eqn:SPDC}  corresponds to $\Delta_+ = 0$. After $p_1$ (or $p_2$) is measured, the standard deviation of $p_2$ (or $p_1$) becomes $\Delta p=\left(\frac{1}{\Delta_+^2}+\frac{1}{\Delta_-^2}\right)^{-1/2}$. 
Similarly, after measuring either position, the standard deviation  of the unmeasured position becomes $\Delta x =  \left(\Delta_+^2 + \Delta_-^2 \right)^{-1/2}$. Thus we arrive at
\begin{equation}
	\Delta x \Delta p = \left(2 + \frac{\Delta_-^2}{\Delta_+^2} +  \frac{\Delta_+^2}{\Delta_-^2} \right)^{-1/2} 
	 \leq \frac{1}{2}
	\,.
	\label{eqn:uncertainty}
\end{equation}
The apparent violation of the uncertainty principle happens whenever $\Delta_+ \neq \Delta_-$. In the extreme case of \equa{eqn:SPDC}, we have  $\Delta x \Delta p =0$.

If momentum is conserved in the type-A device in Fig.~\ref{fig:pn}, we have $p_1 = -p_2$ exactly satisfied. However, the length $L$ of the device limits momentum conservation and leads to $\Delta_+ \sim 1/L$. The damping rate limits the plasmon line width and renders $\Delta_- \sim 1/l$ where $l=v_0/\gamma$ is the propagation length of the mode-0 plasmons. For device satisfying $L \ll l$, we have $\Delta x \Delta p \sim   L/l \ll  1/2$. 
Note that in our case, entanglement is between continuous variables, and so the fidelity of the state can not be defined in the conventional way~\cite{Martin.2017}. Instead, it is more natural to use \equa{eqn:uncertainty} to quantify the entanglement. Alternatively, one could use entanglement witnesses or entanglement negativity~\cite{Pan.2012} which  are more technically involved.

\label{sec:Detection}
\begin{figure}
	\includegraphics[width=0.7 \linewidth]{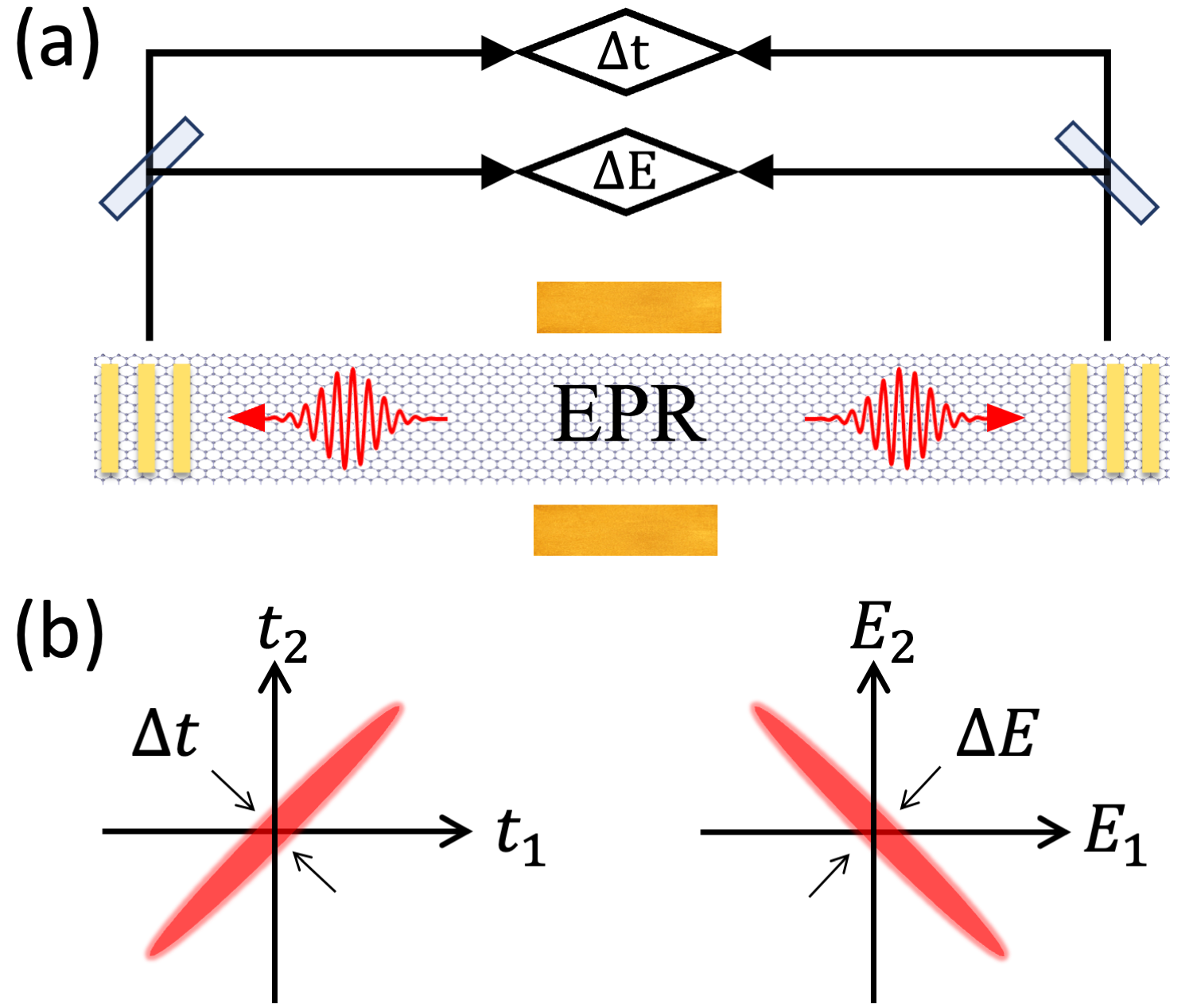} 
	\caption{(a) Schematic of the experiment setup for  coincidence measurement of the energy-time entanglement of the generated EPR plasmon pairs. The plasmon pair generator can be either a type-A (Fig.~\ref{fig:pn}) or type-B device (Fig.~\ref{fig:Current_carrying_device}). The vertical yellow bars are the grating structures to convert plasmons into far field photons. (b) Illustration of the time (left) and energy (right) correlations. The widths of the ellipses are the $\Delta t$ (left) and $\Delta E$ (right). }
	\label{fig:EPR}
\end{figure}

\begin{figure}
	\includegraphics[width=0.7 \linewidth]{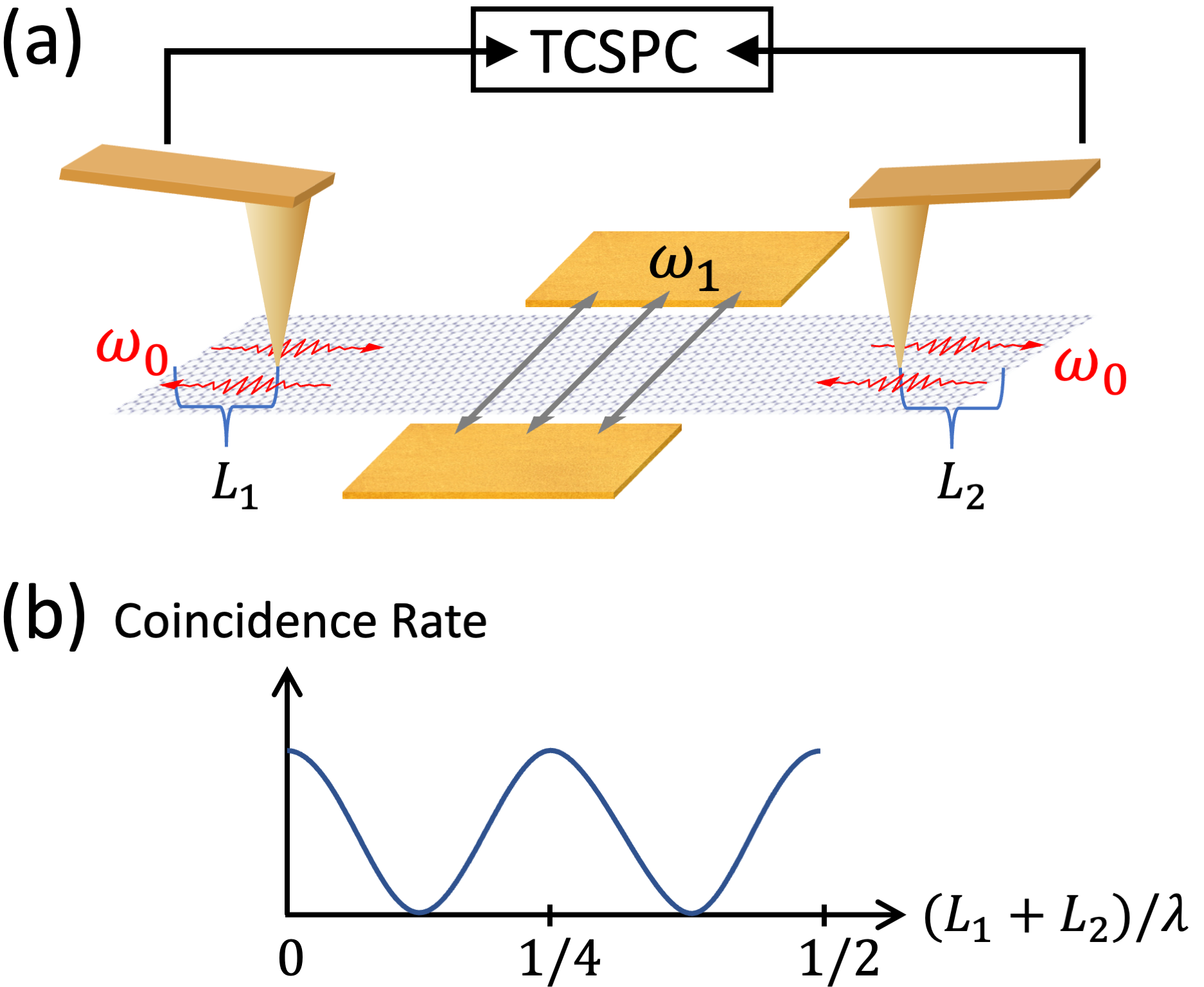} 
	\caption{(a) The Franson scheme of measuring EPR entanglement with two scanning probes as the two interferometers. The plasmon pair generator can be either a type-A  (Fig.~\ref{fig:pn}) or type-B device (Fig.~\ref{fig:Current_carrying_device}). (b) The coincidence rate as a function of  $(L_1+L_2)/\lambda$.}
	\label{fig:SNOM_Franson}
\end{figure}

\begin{figure}
	\includegraphics[width=0.9 \linewidth]{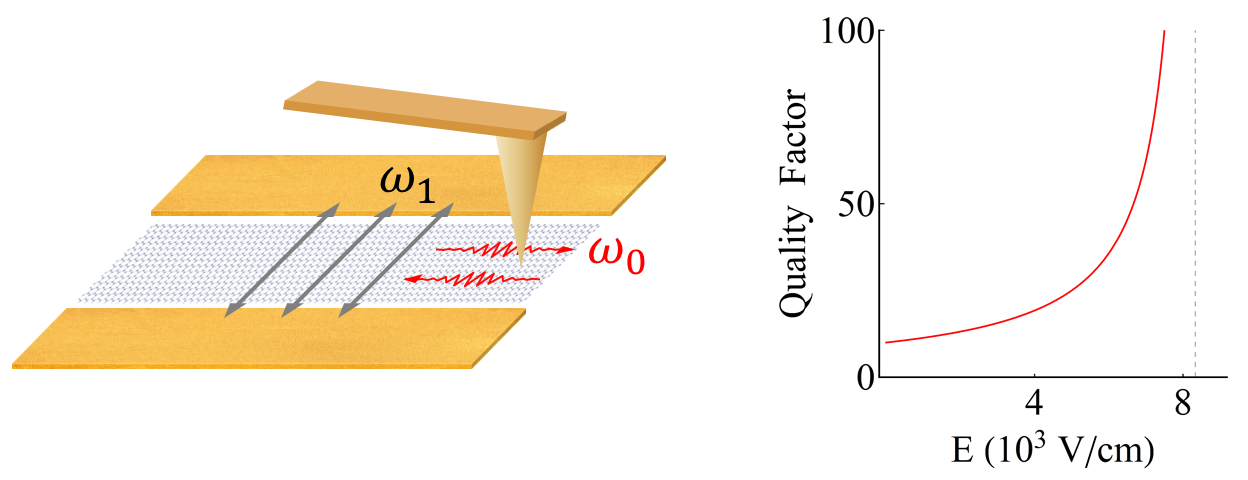} 
	\caption{Near-field imaging of the parametric amplification of the mode-$0$ plasmons. The right figure is plotted for a type-A device using the parameters in Fig.~\ref{fig:pn}. The parametric plasmon amplifier can also be a type-B device in Fig.~\ref{fig:Current_carrying_device}. }
	\label{fig:SNOM_q} 
\end{figure}

To detect the position-momentum entanglement of a plasmon pair, one can place grating structures on both sides of the device which would convert the plasmons to far-field photons.
Passing through the beam splitters, the photons are read either by single photon detectors~\cite{Reid2009, Lee:2020uz} which have good time precision or those with good energy resolution, as shown in Fig.~\ref{fig:EPR}. The coincidence rate measured by the former as a function of path length difference gives a time uncertainty $\Delta t$, while the spectrum correlation of the energy detectors gives the energy uncertainty $\Delta E$. The EPR entanglement is manifested in the relation $\Delta t \Delta E = \Delta x \Delta p < 1/2$ \cite{Reid2009, Boyd.2004,Maclean.2018}. This `energy time entanglement' detection scheme is the same as that in Ref.~\cite{Maclean.2018} with the entangled photon source replaced by the plasmonic type-A device in Fig.~\ref{fig:pn} or type-B device in Fig.~\ref{fig:Current_carrying_device}. Alternatively, homodyne detection experiments can measure the entanglement property of the squeezed state~\cite{Reid2009}. 
	
In addition, the Franson scheme~\cite{Franson.1989,Steiner2021uep, Javid.2021} is also applicable as shown in Fig.~\ref{fig:SNOM_Franson}. As mentioned in \secref{sec:Introduction}, near field technique based on scanning probes has been successful in probing plasmons in graphene~\cite{Basov2014,Fei2012, Chen2012, Woessner2015, Shi:2015vq,Ni2016, Lundeberg2017,Ni2018, Wang:2020vs, Dong:2021tn, Zhao:2021vn, Wang:2021ud}. To implement the Franson scheme, one needs two scanning probes, as shown in \fig{fig:SNOM_Franson}(a).
The plasmons can either travel to the scanning probes directly or after being reflected by the edges of the ribbon, which we call short and long paths respectively. The coincidence detection arises from interference between the short-short and long-long paths. If the coincidence rate between the signals from two near field tips is measured as a function of $L_1 \approx L_2$, the distance between each tip and the edge, it will oscillate as $R_c \sim \cos^2((L_1+L_2)/\lambda) $, like that in Fig.~\ref{fig:SNOM_Franson}(b). However, the condition $L \ll L_i \ll l$ needs to be satisfied. The first inequality guarantees that there are no single photon (or `short-long' path) interference contributions, and the second inequality ensures that plasmon damping is not important. 
Note that in a prior work on measuring plasmon entanglement~\cite{Fasel.2005}, the interfering particles were actually photons in optical fibers. In the proposed scheme (\fig{fig:SNOM_Franson}) the interference occurs between plasmons. Additionally, our implementation involves highly confined terahertz-infrared plasmons and scanning probes operating on deeply subdiffractional length scales.

One can  also probe the classical effects. If the pump field is comparable to the parametric instability threshold, the enhancement of plasmon lifetime can be detected.
Upon pumping of the $\omega_1$ mode of the type-A device in Fig.~\ref{fig:pn} (or type-B in Fig.~\ref{fig:Current_carrying_device}), the $\omega_0$ modes with momenta $q$ and $-q$ are parametrically driven such that their effective damping rate is reduced to $\gamma-\kappa$ (see Appendix~\ref{sec:OPA}), and their quality factor is boosted to $\omega_0/(\gamma-\kappa)$, as shown in Fig.~\ref{fig:SNOM_q}. In a scanning near-field  experiment, its manifestation is simply increased propagation length of the plasmons. One can also measure near field DFG using classical interference between  the signal and idler in these experiments, which is discussed in Appendix~\ref{appendix:DFG_near_field}.

\section{Plasmon generation by the third-order nonlinearity}
\label{sec:third_order}
%
In this section we study the effects of third order nonlinearity~\cite{Mikhailov2007,Soavi:2018aa,Jiang:2018aa, Cheng2014, Mikhailov2017, Sun2018b} on plasmon interactions which occurs already in the long wavelength limit.  This nonlinearity originates from the linearity of the dispersion of Dirac quasiparticles in graphene which breaks Galilean invariance~\cite{Sun2018b}. 
At low frequency $\omega \ll \varepsilon_F$, inter-band effects can be neglected and the third order nonlinear conductivity $\sigma^{(3)}$ in graphene assumes the third order `Drude' form 
\begin{align}
	\sigma^{(3)}_{ilmn} &=\frac{i D^{(3)}}{\omega_1 \omega_2 \omega_3} \Delta_{ilmn} 
	\,
	\label{eqn:sigma_3}
\end{align}
both in the kinetic~\cite{Mikhailov2016} and hydrodynamic~\cite{Sun2018b} regimes where $\omega_i$ are the frequencies of the three electric fields that generate the third order current together. The symmetric tensor $\Delta_{ilmn}$ is the sum of all isotropic tensors $\Delta_{ilmn}=\delta_{il}\delta_{mn}+\delta_{im}\delta_{ln}+\delta_{in}\delta_{lm}$. Close to zero temperature ($T\ll \mu$), the third order optical weight is predicted to be $D^{(3)}=\frac{1}{24\pi} \frac{e^4 v_F}{\hbar^3 k_F}$ in the kinetic regime and twice as large in the hydrodynamic regime~\cite{Sun2018b}. In what follows, we assume the system is the kinetic regime since we are mostly interested in the mid infrared plasmons whose frequencies are larger than the typical electron-electron scattering rate \cite{Sun2018a,Sun2018b}. 

\subsection{FWM Hamiltonian}
We start with the plasmonic effective Lagrangian of an electronic system whose linear conductivity is in the dissipation-less Drude form: $\sigma_{jk} =\frac{i D/\pi}{\omega} \delta_{jk}$ where $D$ is the Drude weight. In the near field approximation (meaning there is only longitudinal electric field and no magnetic field) and in the gauge $A^\mu=(0,\mathbf{A})$ for the electric field where $\mathbf{A}$ is the vector potential viewed as the plasmonic field, the  Lagrangian is
\begin{align}
L &= \int d^3r \left( \frac{1}{8\pi c^2} \dot{\mathbf{A}}^2 - \frac{D}{2\pi c^2} \mathbf{A}^2 \delta(z) \right) + L^{(3)} \notag\\
&=\sum_q \left( \frac{1}{8\pi c^2} \frac{2}{|q|}  \dot{A}_{-q} \dot{A}_q - \frac{D}{2\pi c^2} A_{-q} A_q \right) + L^{(3)}
\,
\label{eqn:l_2D_linear}
\end{align}
where $c$ is the speed of light. The first term is the electric field energy and the second term has the interpretation of the center of mass kinetic energy of the charge carriers. For two dimensional (2D) Drude conductors modeled as the $x$-$y$ plane embedded in 3D space, since the current is localized on the 2D plane, the second term of the top line in Eq.~\eqref{eqn:l_2D_linear} is nonzero only on the plane. 
In the second line of Eq.~\eqref{eqn:l_2D_linear},
$\mathbf{A}_q=\hat{q} A_{q}$ are the Fourier components of $\mathbf{A}$ evaluated on the 2D plane. The $2/|q|$ comes from integrating over the fields exponentially decaying into the three dimensional (3D) space.
Note that the summation is over $\mathbf{q}=(q_x,q_y)$, and instead of three, there is only one plasmon mode for each $\mathbf{q}$ since the field is constrained to be longitudinal. The resulting Euler-Lagrange equation of motion for $A_q$ yields the 2D plasmon dispersion $\omega_q = \sqrt{2Dq}$ where the Drude weight is $D=v_F k_F e^2/\hbar $ for graphene at zero temperature. Quantization of the plasmonic field
\begin{align}
A_q=i\hat{q} A_{qu} (a_q + a_{-q}^\dagger),\quad
A_{qu}/c = \sqrt{\pi \hbar \omega_q/ (2 S D)}
\end{align}
leads to the Hamiltonian
\begin{align}
H=H_0 + H^{(3)},\quad
H_0 = \sum_q \hbar \omega_q \left( a_q^\dagger a_q + \frac{1}{2} \right) 
\label{eqn:H_linear}
\end{align}
where $S$ is the area of the sample.

The $H^{(3)}= -L^{(3)}$ contains products of four plasmonic	 creation and annihilation operators. It describes FWM which is caused by the third order optical conductivity $\sigma^{(3)}$. 
Writing the third order current $j^{(3)}$  in terms of vector potentials,  Eq.~\eqref{eqn:sigma_3} indicates 
$
j^{(3)}_i =\Pi_{ilmn} A^l A^m A^n /c^3  \,
\label{eqn:j3}
$
and the interaction  Hamiltonian 
\begin{align}
H^{(3)} =& - \frac{1}{4c^4} \int d^2r \Pi_{ilmn} A^i A^l A^m A^n 
\notag\\
=& -\frac{D^{(3)}}{4c^4} \sum \Delta_{ilmn} A^i_{q_1} A^l_{q_2} A^m_{q_3} A^n_{q_4} 
\,
\label{eqn:Hc3}
\end{align} 
where the kernel $\Pi_{ilmn}=D^{(3)} \Delta_{ilmn}$ is perfectly local  in space and time, and the summation runs over all $q_i$ constrained by $q_1+q_2+q_3+q_4=0$ due to momentum conservation. Note that although \equa{eqn:Hc3} is a negative $\phi^4$ term in this bosonic field theory, the system is stable due to higher order nonlinear couplings. 

\subsection{Spontaneous four wave mixing}
\label{sec:four_wave_mixing}
\begin{figure}[t]
	\includegraphics[width=0.6 \linewidth]{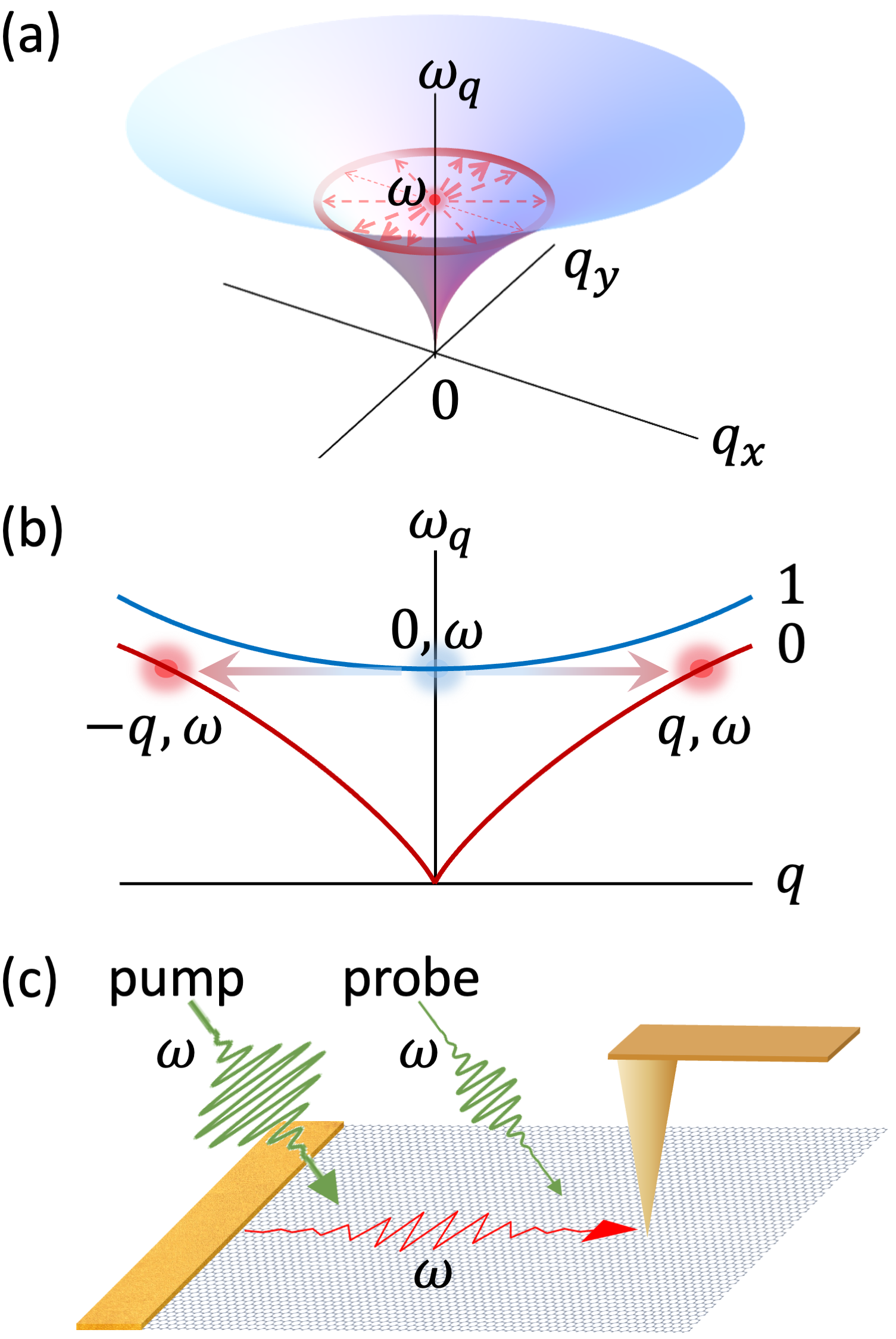} 
	\caption{(a) Generation of entangled plasmon pairs due to spontaneous four wave mixing. The red dot at the center represents the pump. The red circle on the plasmon disperion curve represents the generated plasmons at the same frequency as the pump but nonzero momenta. The opacity of the circle and thickness of the arrows show the angular dependence of the pair generation intensity. (b) The decay of two mode-1 plasmons into a pair of entangled mode-0 plasmons on a graphene ribbon (same as type-A device in \fig{fig:pn} but without doping gradient) due to spontaneous FWM. (c) The schematic of a near field measurement of the enhanced plasmon lifetime on a type-C device.}
	\label{fig:sigma_3_effect}
\end{figure}
This subsection discusses spontaneous FWM shown in Fig.~\ref{fig:sigma_3_effect} in uniformly doped graphene. \equa{eqn:Hc3} implies  that entangled plasmon pairs with frequency $\omega$ can be generated by incident light of the same frequency.
We model far field photon incident on the graphene plane by an uniform AC electric field with amplitude $E_0=\omega A_0$ in $\hat{y}$ direction, and represent it by the vector potential
$
\mathbf{A}_0 = \hat{y} A_0 (e^{-i\omega t} + e^{i\omega t})
\,
\label{eqn:A_0}
$
where we have assumed $A_0$ to be real without loss of generality.
When two fields in the product are replaced by the source $\mathbf{A}_0$, \equa{eqn:Hc3}  becomes
\begin{align}
H^{(3)} 
&= -\sum_{q_x>0} g^{(3)}_{q}(t) (a_q+a_{-q}^\dagger) (a_{-q}+a_q^\dagger)
\,
\label{eqn:Hc3_quantized}
\end{align} 
where the pair generation strength is 
\begin{align}
g^{(3)}_{q}(t) 
&= \frac{1}{8} \hbar \omega_q \xi_\omega^2 \left(1+2\sin^2 \theta_q \right)  \left( 2+e^{-2i\omega t}+e^{2i\omega t} \right)
\,
\label{eqn:gq}
\end{align} 
and $\theta_q$ is the angle of the plasmon momentum $q$ relative to the $\hat{x}$ axis.
We have defined the dimensionless small parameter
\begin{align}
\xi_\omega=\frac{e E_0 /\omega }{ \hbar  k_F } \,
\label{eqn:xi_omega}
\end{align}
which controls the strength of FWM and other third order effects in graphene.
In Eqs.~\eqref{eqn:H_linear} and \eqref{eqn:Hc3_quantized}, it is enough to consider terms close to resonance in the interaction picture. The number conserving terms in Eq.~\eqref{eqn:Hc3_quantized} like $a_q a_q^\dagger$ imply the field induced Kerr shift~\cite{Boyd.2008} of the plasmon frequency, which are red shifts in graphene due to its negative third order conductivity \cite{Sun2018b}. The terms like $e^{-2i\omega t} a_q^\dagger a_{-q}^\dagger$ lead to pair generation in the weak interaction regime ($g^{(3)}Q/\omega \ll 1$ where $Q=\omega/\gamma$ is the plasmon quality factor) and two mode squeezing (instability) in the strong interaction regime ($g^{(3)}Q/\omega > 1$).

In the weak interaction regime, the plasmons with exactly the same frequency as $\omega$ are generated as entangled pairs and with an angular distribution of $\left(1+2\sin^2 \theta_q \right)^2$, as shown in Fig.~\ref{fig:sigma_3_effect}(a).  The pair generation rate is determined by Fermi's golden rule (analogous to \equa{eqn:decay_rate}):
\begin{align}
\Gamma = \frac{2\pi}{\hbar} \langle |g^{(3)}|^2 \rangle_\theta \cdot \frac{S}{16\pi} \frac{\omega^3}{D^2}   \,
\label{eqn:third_order_generation_rate}
\end{align}
where $\langle \rangle_\theta$ means angular average, the second part comes from the 2D  plasmon density of states and we have assumed $\omega \gg T$ so the relevant plasmons are not occupied in equilibrium. 
The resulting dimensionless generation rate is
\begin{align}
\frac{\Gamma}{\omega} = \frac{9}{2^{10}} \xi^4 S \frac{\omega^4}{\alpha_g^2 v_F^2 \omega_F^2}  \,.
\label{eqn:relative_generation_rate}
\end{align}
For $n=10^{12} \unit{cm}^{-2}$, $E_0=10^3 \unit{V/cm}$ and $S=1 \unit{\mu m}^2$, we have $\Gamma/\omega = 3.2 \times 10^{-8}$. This leads to superior generation rates compared with conventional FWM sources,  as we will discuss in Sec.~\ref{sec:Discussion}. Since there is no need to break the inversion symmetry, this effect could be observed simply using uniformly doped graphene, which we call the type-C device. One may also use the ribbon in Fig.~\ref{fig:pn}(a) but a lateral $p$-$n$ junction is no longer necessary. The spontaneous FWM process in a transversely pumped ribbon is illustrated in \fig{fig:sigma_3_effect}(b). To distinguish the generated entangled plasmon pairs from the linearly excited plasmons, the key experimental signature would be satisfaction of the EPR criterion~\cite{Reid2009} in the quadrature measurement in Fig.~\ref{fig:EPR} or the nonlocal interference phenomenon in the Franson experiment in Fig.~\ref{fig:SNOM_Franson}.

In the strong interaction regime, the plasmons are squeezed in pairs and experience parametric amplification. The resulting relative exponential growth rate is (similar to \equa{eqn:Qg})  
\begin{align}
	Q_g^{-1}= \frac{|g^{(3)}_{q}|}{\hbar \omega}  = \frac{1}{8} \xi^2 \left(1+2\sin^2 \theta_q \right) \,
	\label{eqn:relative_growth_rate_third}
\end{align}
for the the plasmons propagating in $\theta_q$ direction, which agrees with the classical result (see Appendix~\ref{sec:third_order_apd}). In a typical near field experiment shown in Fig.~\ref{fig:sigma_3_effect}(b), this  reduces the plasmon damping rate from $\gamma$ to $\gamma-g_q^{(3)}/\hbar$. Therefore, to overcome the typical damping rate  $\gamma \sim 0.1 \omega $ of graphene plasmons,  one  needs $\xi \sim 0.3$ which requires a pump field of $E_0\sim 10^5 \unit{V/cm}$ for $n=10^{12} \unit{cm}^{-2}$ and $\omega =30 \unit{THz}$.

\subsection{Spontanenous PDC in current carrying graphene}
\label{sec:spdc_current}
\begin{figure}[t]
	\includegraphics[width=0.7 \linewidth]{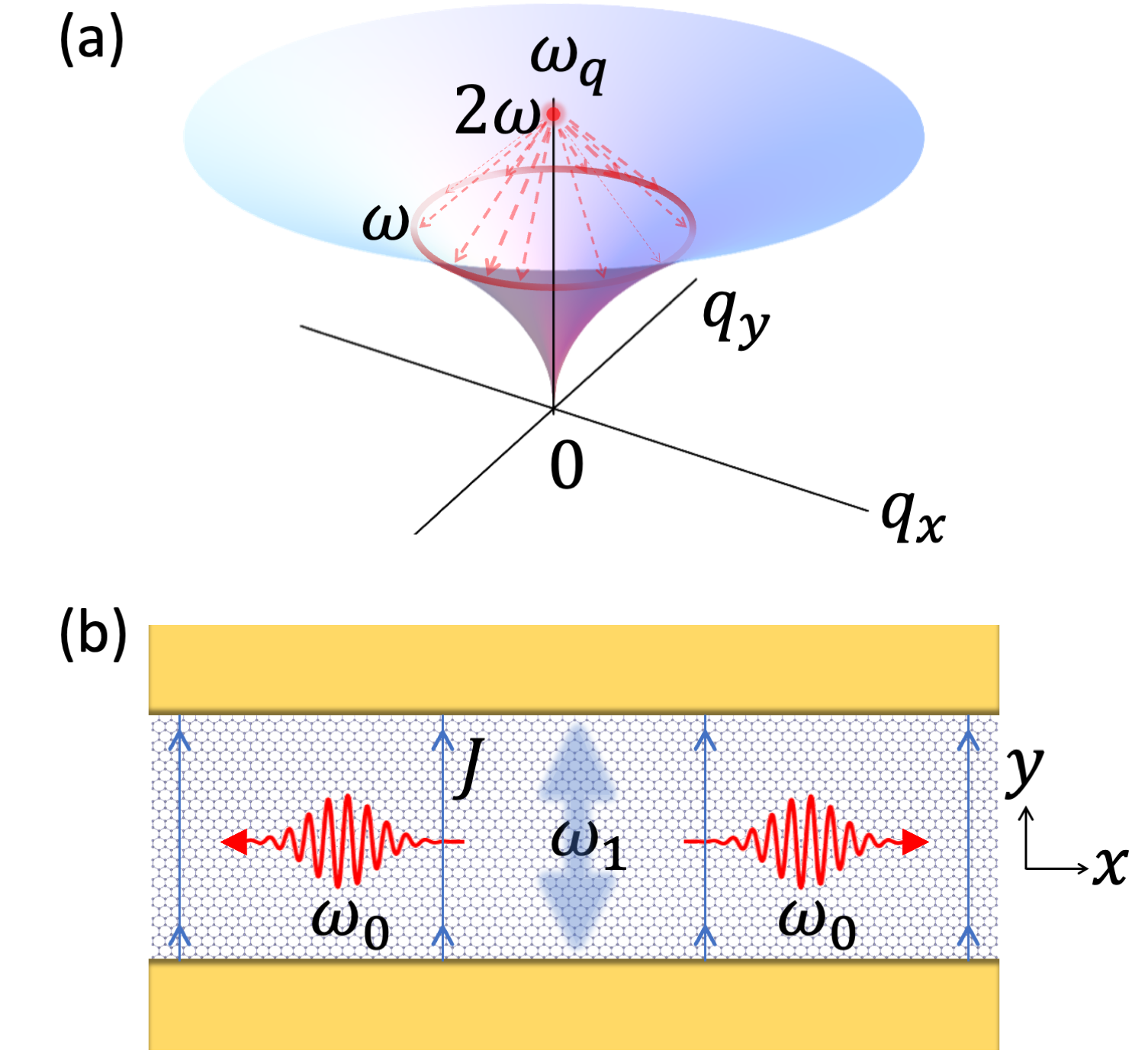} 
	\caption{(a) Parametric down conversion of the pump of frequency $2\omega$ into pairs of subharmonic plasmons of frequency $\omega$ in current carrying graphene. (b) Realization in a ribbon (type-B device) which is similar to the type-A device in Fig.~\ref{fig:pn} except that the density gradient is replaced by a dc current $J$ flowing between the contacts.}
	\label{fig:Current_carrying_device}
\end{figure}
The third order nonlinearity can also lead to PDC in a current-carrying graphene, as shown in Fig.~\ref{fig:Current_carrying_device}. This effect can also be understood as FWM with one ``pump wave'' in \equa{eqn:Hc3}  replaced by a dc current flow ($\omega=0$). If the other pump source has frequency $2\omega$, then the process looks like a PDC: $2\omega\rightarrow \omega + \omega$. Another point of view is that the dc current  with flow velocity $u$ breaks the inversion symmetry, resulting in a nonzero second order nonlinear conductivity $\sigma^{(2)}\sim \frac{u}{v_F} \frac{e^3 v_F}{\hbar^2} \frac{1}{\omega^2}$. To leading order in the dc current $J=enu$, the equations from subsection~\ref{sec:four_wave_mixing} can be directly applied to the present case. We assume that the pump field is polarized parallel to the dc current, such that the PDC Hamiltonian is the same as Eqs.~\eqref{eqn:Hc3_quantized} and \eqref{eqn:gq} but with $\xi_\omega^2$ replaced by $\xi_{2\omega} \xi_{0}$ where $\xi_{0}=eE_{\text{dc}}/(\gamma_{\text{dc}} \hbar k_F)\approx u/v_F$ is the dimensionless parameter due to the dc electric field $E_{\text{dc}}$ that drives the dc current.

In the weak pump regime, entangled  subharmonic plasmon pairs at frequency $\omega$ are generated with the angular distribution of $\left(1+2\sin^2 \theta_q \right)^2$ in an infinite graphene sheet, same as that of spontaneous FWM.
Summed over all angles, the total relative  generation rate is the same as \equa{eqn:relative_generation_rate} but with  with $\xi^4$ replaced by $\xi_{2\omega}^2 \xi_{0}^2$. 
To measure the entanglement property of the emitted pairs, one can use the graphene ribbon in Fig.~\ref{fig:Current_carrying_device}(b) as a plasmon pair source to perform either the energy-time entanglement measurement in Fig.~\ref{fig:EPR} or the Franson scheme in Fig.~\ref{fig:SNOM_Franson}. In the ribbon, the generated subharmonic pairs are not in every direction, but can only propagate along $\hat{x}$ (similar to  Fig.~\ref{fig:pn}(b)) with the generation rate described by roughly the same formula as that for an infinite sheet.
For $n=10^{12} \unit{cm}^{-2}$, $\omega_1=2\omega = 30 \unit{THz}$, $E_0=10^3 \unit{V/cm}$, a ribbon size of $S= 0.2\unit{\mu m} \times 5\unit{\mu m}$ and a dc current of $J= 0.16 \unit{mA/\mu m}$ corresponding to $u=0.1 v_F$, one has  $\xi_{0} \approx 0.1$ and a normalized generation rate of  $\Gamma/\omega \approx 0.9 \times 10^{-6}$ from this type-B source. 

In the strong pump regime, the  plasmon lifetime is enhanced.
The  pump induced relative growth rate of plasmons of frequency $\omega$ is  
\begin{align}
Q_g^{-1}= \frac{1}{16} \xi_{2\omega} \xi_{0}  \Delta_{ilmn} \hat{q}_i \hat{q}_l \hat{E}_m \hat{J}_n \,,
\label{eqn:relative_generation_rate_current}
\end{align}
similar to \equa{eqn:relative_growth_rate_third}. Here $\hat{X}$ means the unit vector along $X$ and $\hat{\mathbf{q}}$ is the unit vector of the momentum of the subharmonic plasmon. For a current of $J=0.5 \unit{mA/\mu m}$ at the doping level of $n=10^{12} \unit{cm^{-2}}$ and under a pump field of $3 \times 10^5 \unit{V/m}$ at $\omega_1=2\omega = 30 \unit{THz}$ polarized along $\hat{y}$, the subharmonic plasmon running along $\hat{y}$ has a relative growth rate of $Q_g^{-1} \approx 0.03$, while those along $\hat{x}$  has a growth rate of $Q_g^{-1} \approx 0.01$, comparable to the plasmon loss at low temperatures~\cite{Ni2018}.

\subsection{Field enhancement in a ribbon}
From previous discussion, high field is essential to make the proposed devices practical. By design, these devices naturally enhance the incident field by two mechanisms, helping to achieve this goal.
First, as discussed in Sec.~\ref{sec:Two_mode_squeezing}(a), the split gate for type-A device in \fig{fig:pn}, or the contacts for type-B device in Fig.~\ref{fig:Current_carrying_device}(b), act as antennas. They enhance the pump field by a factor of $F=E/E_v= \lambda_v/(2\pi w_s) \approx 10$~\cite{Jiang:18} for the channel width $w_s=200\unit{nm}$ and the vacuum wave length of the pump light  $\lambda_v \approx 10 \unit{\mu m}$ at $\omega_1=30 \unit{THz}$. 
Second, close to the resonance at $\omega_1$ with the gapped mode of the ribbon (see, e.g., Fig.~\ref{fig:pn}(b) and \fig{fig:sigma_3_effect}(b)), there is another field enhancement factor of $F_Q=\xi_f \frac{\omega_1^2}{(\omega-\omega_1+i\gamma)(\omega+\omega_1+i\gamma)}$ where $\xi_f$ is an $O(1)$ shape factor, see Eqs.~\eqref{eqn:EOM_source} and \eqref{eqn:efield_profile}.  On resonance, this leads to a field enhancement factor of $F_Q \sim Q \sim 10$.

The antenna effect together with  the resonance field enhancement gives a net enhancement factor of $QF \sim 100$. Therefore, the spontaneous PDC rate in the type-B device (Fig.~\ref{fig:Current_carrying_device}) is enhanced by a factor of $(QF)^2$ while the spontaneous FWM rate in a ribbon is enhanced by $(QF)^4$.
Similarly, to reach a pump field of $10^5 \unit{V/m}$ for parametric instability, the external incident field just needs to be of the order of $10^3 \unit{V/m}$.

Note that the resonant field enhancement also occurs in photonic cavities such as micro-ring resonators~\cite{Absil2000,Helt2010,Ferrera:2008wi,Levy:2010vn,Steiner2021uep}. In our devices, the vacuum photons play the role of the waveguide photons in the bus waveguide therein, and the graphene ribbon plays the role of the ring resonator. Unlike the plasmon modes in the nano-ribbon whose line width is dominated by intrinsic damping, the field enhancement on resonance in the micro-rings  scales as $\sqrt{Q}$ instead of $Q$ \cite{Helt2010} because the line width $\gamma_p$ of the photon modes in the micro-ring resonator is dominated by radiation leakage into the bus waveguide. As a result, the 
enhancement of the FWM pair generation rate due to resonant pumping scales as $Q^2$, see, e.g., Eq.~4 of Ref.~\cite{Absil2000}. 
Note also that in both our devices (short devices with length $L_0\ll l$) and photonic cavities, the resonance conditions with the generated modes contributes another factor of $Q$ to the enhancement of the generation rate. 

\section{Discussion and experimental outlook}
\label{sec:Discussion}

\begin{table*}
	\begin{ruledtabular}
		\begin{tabular}{llccc}
			PDC Devices & Frequency & $Q$ &  $\chi^{(2)} $ ($\unit{cm/V}$) & Generation Rate ($\unit{s^{-1} mW^{-1}}$)
			\\
			\hline 
			Graphene type-A device (Fig.~\ref{fig:pn}(a)) & $4 \unit{THz}$  & $10 - 100$  & $ 2\times 10^{-6} $ &  $9 \times 10^{11} $
			\\[3pt]
			Graphene type-B device (Fig.~\ref{fig:Current_carrying_device}(b)) & $15 \unit{THz}$  & $10-  100$  &  $ 3\times 10^{-6} $ &  $2 \times 10^{12} $
			\\[3pt]
			Semiconductor coupled quantum wells~\cite{sirtori1994, Ding.1995, Belkin2007}  & $1  - 50 \unit{THz}$ & ---  & $ \sim 10^{-6}$ &  ---  
			\\[3pt]
			LiNbO$_3$ \cite{Zhao.2020}  & $0.79 \unit{eV}$ & ---  & $ 2 \times 10^{-10}$ &  $2 \times 10^{7} $
			\\[3pt]
			AlN \cite{Guo:2017ue}  & $0.80 \unit{eV}$ &  $1 \times10^{5}$ & $ 4 \times 10^{-11} $  &  $6 \times 10^{6} $
			\\
			\hline
			FWM Devices & Frequency &$Q$ & $\chi^{(3)} $  ($\unit{cm^2/V^2}$) & Generation Rate ($\unit{s^{-1} mW^{-2}}$)
			\\
			\hline
			Graphene type-B device with no DC current & $30 \unit{THz}$ & $10 - 100$ & 
			$ 4 \times 10^{-14} $  
			& $ 6 \times 10^{14} $
			\\[3pt]
			AlGaAs on Insulator ~\cite{Steiner2021uep} & $0.80 \unit{eV}$  & $1 \times 10^6$ &
			$  7 \times 10^{-17} $   & $2 \times 10^{10} $
			\\[3pt]
			Silicon on Insulator \cite{Koos2007,Ma2017} & $0.80 \unit{eV}$ & $\sim 10^5$  &
			$  4 \times 10^{-17} $  &  $2 \times 10^{8} $
			\\[3pt]
			InP \cite{Kumar2019}  & $0.83 \unit{eV}$  & $4 \times 10^4$  & $2 \times 10^{-16} $ &  $1 \times 10^{8} $
			\\[3pt]
			Si$_3$N$_4$  \cite{ramelow2015} &$0.80 \unit{eV}$  & $2 \times 10^6$  & $4 \times  10^{-19} $   &  $4 \times 10^{6} $
		\end{tabular}
	\end{ruledtabular}
	\caption{ Comparison of graphene with conventional platforms for entangled photon generation following that in Ref.~\cite{Steiner2021uep}. Although plasmons in these graphene devices can work in a broad frequency range from terahertz to infrared, we picked specific frequencies for estimations in the table.  For graphene type-A device, the plasmon frequency is shown in Fig.~\ref{fig:pn}(b) while for type-B device (with a current flow velocity of $u=0.1 v_F$) and FWM, we used a pump frequency of $30 \unit{THz}$. We used $a=200 \unit{nm}$ and the quality factor of $Q=10$ for all the graphene devices.   Since we are using the Gaussian unit system while the majority of nonlinear optics literature uses the SI unit, we translated from the latter in the literature to the former in this table.   The refraction indices used for conversion of $\chi^{(3)}$  are AlGaAs: $N=3.2$ at $\lambda=1.55 \unit{\mu m}$, Silicon: $N=3.5$ at $\lambda=1.55 \unit{\mu m}$, InP: $N=3.4$ at $\lambda=1.49 \unit{\mu m}$,  Si$_3$N$_4$: $N=2.0$ at $\lambda=1.55 \unit{\mu m}$.}
	\label{tbl:chi_chi}
\end{table*}

We proposed several graphene  spontaneous PDC  and FWM devices which have  promising quantum information applications~\cite{Alonso-Calafell.2019} as efficient sources of entangled plasmon pairs, photon-plasmon  converters and plasmon amplifiers~\cite{Page2015, Sun2016, DeVega2017, Low2018, Rajasekaran2016a}, etc., for future quantum plasmonic/polaritonic circuits.  
These plasmons are in the THz-Infrared spectra range, which is an important energy range in condensed matter physics where superconductivity, magnetism, spin liquid, fractional quantum hall effect and other correlated phenomena are relevant. The quantum sources proposed here may enable quantum imaging~\cite{Moreau:2019uk} and sensing \cite{Pirandola:2018vs} of these novel states of matter.

Let us compare our devices with conventional photonic PDC sources. In photonic PDC,
the nonlinearity is usually characterized by a nonlinear susceptibility $\chi^{(2)}$,
which has the unit of inverse electric field.
We can convert the 2D conductivity $\sigma^{(2)}$ to an effective 
3D nonlinear susceptibility $\chi^{(2)}$
as follows.
The 2D nonlinear susceptibility (in the Gaussian unit system) is~\cite{Yao2014} $\chi^{(2)}_\mathrm{2D} =  i \sigma^{(2)} / \omega_0$.
For the type-A device which is a ribbon with half width $a$, the extent of the electric field of the plasmon in the out-of-plane direction is $1 / q_0 \sim a$. Hence, the effective 3D nonlinear susceptibility is $\chi^{(2)} \sim \chi^{(2)}_\mathrm{2D} / a  \sim 1 / (e k_F^2)$. 
For $k_F = \sqrt{\pi} \times 10^6 \unit{cm^{-1}}$ corresponding to carrier density $n = 10^{12} \unit{cm}^{-2}$, we find $\chi^{(2)} \sim 2\times 10^{-6} \unit{cm/V}$,
which is much larger than that of conventional photonic media, e.g., LiNbO$_3$~\cite{Zhao.2020,Javid.2021}, but is
of the same order as $\chi^{(2)}$ in sophisticated  photonic waveguides made of coupled quantum wells ~\cite{sirtori1994, Ding.1995, Belkin2007}, see Table~\ref{tbl:chi_chi}.  Nevertheless, a more important figure of merit is the dimensionless decay factor 
\begin{align}
	\frac{\Gamma}{\omega_0} \sim |\chi^{(2)}|^2 \frac{\hbar\omega_0}{a^3}
	\label{eqn:gamma2}
\end{align}
(same as \equa{eqn:decay_factor} noting that $g^{(2)} \sim \chi^{(2)} \omega_0^{3/2}a^{-1} L^{-1/2}$ )
of a waveguide mode into subharmonic modes due to spontanenous PDC.
Due to strong mode confinement ($a \ll c / \omega_0$), this figure of merit is much larger than those of  photonic devices.
Physically, this is because electric field of a single plasmon in our device is much larger than that of a single photon. 

As discussed in Sec.~\ref{sec:spdc_current}, a comparable $\chi^{(2)}$ can also be achieved in a type-B device where inversion symmetry is broken by applying a DC current flow to an uniformly doped ribbon if the flow velocity is moderately fast: $u > 0.1 v_F$. 

We also consider  the FWM~\cite{Boyd.2008} in Sec.~\ref{sec:four_wave_mixing}, a third-order 
nonlinear effect. Such a process does not require inversion-symmetry breaking and can take place in either a ribbon or a plain graphene sheet (type-C device in \fig{fig:sigma_3_effect}). This plasmon instability requires a relatively strong incident field controlled by the small parameter $\chi^{(3)}$, but it may be easier to realize experimentally since there is no specific requirement for the nano structure.   Compared with other nonlinear materials, $\chi^{(3)}=i\sigma^{(3)}/(\omega a)$ in graphene  is much stronger, as shown in Table~\ref{tbl:chi_chi}.

A brief discussion of units conversion is in order.
The Gaussian unit system defines the 3D nonlinear susceptibilities in terms of the effective dielectric function $\epsilon=1+4\pi \left[\chi + \chi^{(2)}E + \chi^{(3)}E^2 + ...\right]$ while the SI system defines them as $\epsilon/\epsilon_0=1+\chi_{\text{SI}}  + \chi^{(2)}_{\text{SI}}E + \chi^{(3)}_{\text{SI}}E^2 + ...$~\cite{Boyd.2008} where  $\epsilon_0$ is the vacuum permittivity. Therefore, besides converting the units of electric fields, there is an additional $4\pi$ factor: $\chi^{(n)}=\chi^{(n)}_{\text{SI}}/(4\pi)$.  The unit $\unit{m^2/W}$ is also used in the literature~\cite{Koos2007,Steiner2021uep} for third order nonlinearity $N_3$ defined as $\delta N =\frac{1}{2N} \chi^{(3)}_{\text{SI}}E^2 =\frac{4\pi}{2N} \chi^{(3)} E^2 \equiv N_3 P$ where $N=\sqrt{1+4\pi \chi}$ is the linear refraction index, $\delta N$ is the change of effective refraction index due to third order nonlinearity and the pump field, and $P$ is the power flux of the pump field. Therefore, besides converting $\unit{m^2/W} \rightarrow 0.00132806 \unit{m^2/V^2}$, one also needs the conversion $\chi^{(3)}=\frac{N}{2\pi} N_3$ to obtain the $\chi^{(3)}$ in Gaussian unit.


We note that the properties of the type-A device may have substantial temperature dependence arising from the middle of the  ribbon [Fig.~\ref{fig:pn}(a)]  where the system is close to charge neutrality. The gapped mode is an anti-symmetric charge oscillation between the upper and lower half of the strip, and thus strongly depends on the conductivity at the middle of the junction where the carriers are thermally excited electrons and holes.  Therefore, the frequency of the gapped mode (the shape factor $\xi_1$ in Sec.~\ref{sec:main_results}) increases with temperature. 
In our derivation, we included only the intraband contributions~\cite{Cheng2016, Wang2016, Tokman2016,Sun2018a} to the second order nonlinear conductivity [\equa{eqn:Pi}]. Close to the middle of the junction where $\mu \lesssim \hbar \omega$, interband contributions may be important.  There the total $\sigma^{(2)}$ is suppressed by a factor of $(\mu/ \hbar\omega)^M$ with $M \geq 2$ compared to the intraband expressions used in this paper,  but $\sigma^{(2)}$  may diverge at the narrow region where $\mu=\hbar \omega $ at the interband threshold for Pauli blocking \cite{Yao2014,Wang2016,Cheng2016,Tokman2016,Rostami2016,Wolff_2019}. This divergence is rounded by nonzero temperature, and would thus lead to substantial temperature dependence if included.	
We note that even in homogeneously doped graphene, there are  $3 \sim 4$ orders of magnitude inconsistencies~\cite{Tollerton.2019} between theories (e.g., Refs.~\cite{Yao2014,Wang2016,Cheng2016,Tokman2016,Rostami2016}) and experiments~\cite{Yao2018,Constant2016}. At this stage, it is premature to include such details in the model, and instead, we provide order of magnitude scaling results for the PDC. Incidentally, experiment and theory seem to agree for $\sigma^{(3)}$~\cite{Jiang:2018aa}.

As mentioned in Sec.~\ref{sec:Introduction}, generation of plasmons by PDC in graphene has also been considered by Ref.~\cite{Tokman2016}. In that study, the pump and idler are far field photons. In contrast, here all three modes are plasmons such that frequencies are much lower and momenta are much larger, making this process much more efficient. Moreover, our proposed device involves an antenna that further increases the coupling efficiency to far field pump. As a result, the threshold power to achieve parametric instability is about four orders of magnitude lower than Ref.~\cite{Tokman2016}. 

Since plasmons do not have the polarization degree of freedom (they are always longitudinally polarized),  entanglement of polarization frequently discussed in quantum optics does not apply here. In plasmonics one focuses on the position-momentum (or energy-time) entanglement. For example, it has been shown this entanglement survives  photon-plasmon conversion processes~\cite{Tame2013,Altewischer:2002aa,Fasel.2005}. Here we proposed a scheme to directly generate and measure the entanglement of plasmon pairs  on a chip using the modern tools of near field optics.

The following estimate shows that graphene will not be damaged by the strong electric field required. The incident light heats up the electron gas in graphene with the heating power $P_{\text{heat}} = 2 \mathrm{Re}[\sigma] |E|^2$ and electron phonon scattering provides its cooling channel. The cooling power is approximated by $P_{\text{cool}} = C_v \gamma_{c} (T_e - T_l)$ where $T_e$($T_l$) is the electron(lattice) temperature, $C_v \sim  g(\varepsilon_F)  T$ is the heat capacity of the electron gas and $\gamma_c \sim 1 \unit{ps}^{-1}$ is the cooling rate revealed by previous experiment~\cite{Ni2016}. The lattice is a good heat conductor and is assumed to stay at the same temperature as the environment.  The balancing of heating and cooling $P_{\text{heat}} = P_{\text{cool}}$ determines the stationary electron temperature $T_e -T_l \sim \frac{1}{4\pi^2}  \frac{\gamma}{\gamma_c} \frac{\varepsilon_F^2}{T} \xi^2 $ where we have set the Boltzmann constant to one. At the typical doping level and frequency scale considered in this paper, for $T_l=300 \unit{K}$ and $E=10^4 \unit{V/cm}$, we conclude that $T_e -T_l \sim 6 \unit{K}$ which is quite small compared to either room temperature or the Fermi energy.

We worked in the kinetic regime where plasmon frequency is much larger than the electron-electron scattering rate  $\omega \gg \Gamma_{ee}$~\cite{Sun2018a}. 
In the low frequency hydrodynamic regime $\omega \ll\Gamma_{ee}$ \cite{Phan2013,Sun2016, Sun2018a,Svintsov.2018, Lucas.2018,Patrick.2019,Andreeva.2020,Narozhny.2021,Levchenko.2020}, the nonlinear conductivities are similar in magnitude as long as the temperature is not much larger than chemical potential~\cite{Sun2018a,Sun2018b, Principi.2019}. Therefore, our estimate of the generation rates should apply as well to the collective modes in the hydrodynamic regime such as the charged `demons'~\cite{Phan2013,Sun2016, Sun2018a,Svintsov.2018, Lucas.2018,Andreeva.2020,Narozhny.2021}. However, in the case of $T \gg \mu$, the demons become almost charge neutral, driven mainly by kinematic pressure,  and the physics could be quite different. It may be interesting to study  nonlinear and quantum effects for these collective modes \cite{Svintsov.2013,Sun2018b}. 

Since the ribbon width is much larger than the lattice constant, the Dirac dispersion approximation for the electrons in graphene are valid. We used the `local' approximation to compute the plasmonic modes, which neglects  dependence of the optical conductivity $\sigma(\omega,q)$ on the field wave vector $q$, see Appendix~\ref{appendix:nano_structure}. The nonzero $q$ effects enter $\sigma(\omega,q)$ as a power expansion in the dimensionless small number $v_F^2 q^2/\omega^2$~\cite{Sun2018a, Lundeberg2017,Dong:2021tn}, see, e.g., the supporting information of Refs.~\cite{Sun2018a,Dong:2021tn}. For the relevant modes of the ribbons such as mode-1 in Fig.~\ref{fig:pn}, one has $q \approx 2\pi/(4a)$ and $v_F q/\omega\approx0.36$, meaning that the devices are indeed in the local regime. As an estimate, the leading order nonlocal correction $\delta \omega_1$ to the plasmon frequency $\omega_1$ is just $\frac{\delta \omega_1}{\omega_1}=\frac{1}{2}  \frac{3}{4} (\frac{v_F q}{\omega})^2 \approx 0.05$ in the kinetic regime we are considering. 
Therefore, the Drude (or `local') approximation is well justified.

Extension of these nonlinear effects to polaritonic modes~\cite{Basov2016, Low:2017wb,Basov.2021} beyond those in graphene is a meaningful future direction. For example, monolayer hexagonal boron nitride~\cite{Dai.2019,Li:2021tj} should exhibit strong second order optical nonlinearity due to broken inversion symmetry of the crystal lattice, and would be a natural platform for generating	 entanglement between the long lived hyperbolic phonon polaritons \cite{Dai.2014,Yoxall:2015vy,Li:2015wj, Sun.2015,Rivera.2017,Giles2018,Brown.2018,Lee:2020ue,Ni:2021tc,Moore:2021wi}. Similar nonlinear processes exist for optical phonons in SiC~\cite{Cartella12148}, Josephson plasmons in layered superconductors~\cite{Rajasekaran2016a,Michael.2020,Sun.2020_collective,dolgirev2021periodic} and the collective modes in excitonic insulators~\cite{Golez.2020,Sun.2020BS,Murakami.2020,Kaneko.2021,Sun.2021_Josephson}.

\acknowledgments
M.M.F and Z.S. were supported by the Office of Naval Research under grant N00014-18-1-2722. Z.S.  acknowledges support from the Harvard Quantum Initiative through the  Postdoctoral Fellowship in Quantum Science and Engineering.
Research at
Columbia University was supported by the US Department of Energy (DOE), Office of Science, Basic Energy Sciences (BES), under award No. DE-SC0018426.
D.N.B. is an investigator in Quantum Materials funded by the Gordon and Betty Moore Foundation's EPiQS Initiative through Grant No. GBMF4533. We thank Y. Shao for bringing to our attention the current carrying devices, and thank I. L. Aleiner, A. J. Sternbach, S. Moore, R. Jing, Y. Dong, L. Xiong and B. S. Kim for helpful discussions.

\appendix
\section{Summary of notations for the ribbon case}
\label{appendix:notations}

The main results of this paper are written in terms of
various ``shape'' parameters and physical quantities such as size ($a$, $L$), Fermi momentum ($\varepsilon_F$), Fermi velocity ($v_F$),  electric field ($E$) and frequency ($\omega$).
Most of the shape parameters are dimensionless numbers of the order of unity. They are $S_0^\prime$, $S_1^\prime$, $\alpha^\prime$, $\eta^\prime$, $\xi_0$, $\xi_1$, $\xi_g$, $\xi_{\text{df}}$, $\xi_\beta$, and $\xi_Q$. 
Their definitions for the case of a ribbon are as follows:
\begin{align*}
S_0^\prime &= \int_{-1}^{1} dy \tilde{\sigma}(y) |\mathbf{\tilde{E}}_0|^2\,, 
\\
S_1^\prime &= \int_{-1}^{1} dy \tilde{\sigma}(y) |\mathbf{\tilde{E}}_1|^2\,,
\\
\alpha^\prime &= \int_{-1}^{1} dy \tilde{\sigma}(y) \hat{\mathbf{e}} \cdot \mathbf{\tilde{\mathbf{E}}}_1
\,,
\\
\eta^\prime &= q_x^{\prime 2} \int_{-1}^{1} dy \, \partial_y^2 \varphi_1 \,
\varphi_0^2 \, \tilde{g}(y)\,,
\\
\xi_v(q) &= \dfrac{\omega(q)}{\sqrt{\alpha_g (a k_F)} v_F q}
= \dfrac{\omega(q)}{v_a q}\,,
\\
\xi_0 &= \xi_v(q_0)\,,
\\
\xi_1 &= \dfrac{\hbar \omega_1}{\sqrt{2e^2 E_F / a}}\,,
\\
\xi_g &= \xi_1^{1/2} \frac{\eta^{\prime}}{S_1^{\prime 1/2} S_0^\prime}\,,
\\	         
\xi_{\text{df}} &= \frac{v_a}{v_l} \frac{\eta^\prime}{S_1^\prime S_0^{\prime 2}} \,,
\\
\xi_\beta &= \frac{1}{\xi_1^{3/2}}\, \frac{\alpha^\prime}{S_1^{\prime 1/2}} \,,
\\
\xi_Q &= \frac{\eta^\prime}{\xi_1^{2}}\, \frac{\alpha^\prime}{S_0^\prime S_1^\prime}\,.
\end{align*}
The remaining ones, $S$, $\alpha$ and $|\tilde{\rho}|^2$, are just the corresponding dimensionless ones multiplied by $a L$,
one half of the total area of the ribbon.

\section{Parametric oscillator}
\label{sec:OPA}

The spontanenous PDC process in Sec.~\ref{sec:main_results} is similar to what happens in a parametric oscillator (PO) except that the former leads to two-mode squeezing while the latter corresponds to single-mode squeezing~\cite{Caves1985}.
Optical POs are widely used to generate subharmonic light
and to create entangled photon pairs.
To make the paper self-contained, we review the basic
theory of the PO in this Appendix.

A PO can be modeled with the equation of motion
\begin{align}
\hat{L} x(t) =
\left[ \partial^2_t + \gamma \partial_t
 + \left( 1 + 2\delta \cos\Omega t \right) \omega_0^2 \right]
x(t) = F(t) \,,
\label{eqn:OPO}
\end{align}
where $\Omega$ is the frequency of the pump
modulating the natural frequency $\omega_0$ of the oscillator by a relative amount $\sim \delta$.
The difference
\begin{equation}
\Delta = \Omega - 2\omega_0
\label{eqn:Delta}
\end{equation}
of $\Omega$ from the primary parametric resonance
frequency $2\omega_0$ is referred to as the detuning.
Given the initial conditions $x(0)$, $\dot{x}(0)
\equiv \partial_t x(t)|_{t = 0}$,
the solution of Eq.~\eqref{eqn:OPO} can be written in terms of the retarded Green's function $G = G(t, t^\prime)$:
\begin{equation}
\begin{split}
x(t) &= \left.\left\{\left[\dot{x}(0) + \gamma x(0)\right] G
- x(0) \partial_{t^\prime}G \right\}
\right|_{t^\prime = 0}\\
&+ \int G(t, t - \tau) \Theta(t - \tau) F(t - \tau)d \tau\,,
\label{eqn:PO_sol}
\end{split}
\end{equation}
where $\Theta(t)$ is the Heaviside step-function.
One simple example is $x(0) = 0$, $\dot{x}(0) = 1$, $F(t) \equiv 0$,
in which case $x(t) = G(t, 0)$.

As usual, the Green's function is constructed from
an arbitrary pair of linearly independent solutions $x_1(t)$, $x_2(t)$ of
the homogeneous equation $\hat{L} x(t) = 0$:
\begin{align}
G(t,t^\prime) &= \Theta(t-t^\prime) \frac{x_1(t^\prime) x_2(t) - x_2(t^\prime) x_1(t)}{W(t^\prime)}  \,,
\label{eqn:G}\\
W(t) &= x_1(t) \partial_t x_2(t)
- x_2(t) \partial_t x_1(t)\,.
\label{eqn:Wronskian}
\end{align}
By virtue of the Abel formula, the Wronskian determinant [Eq.~\eqref{eqn:Wronskian}] has the following $t$-dependence:
\begin{align}
W(t) = W(0) e^{-\gamma t}\,.
\label{eqn:Wt}
\end{align}
One possible choice of $x_1(t)$ and $x_2(t)$ is
\begin{align}
x_1(t) &= e^{-\frac{1}{2}\gamma t} \mathrm{C}
\left( \frac{4\omega_0^2-\gamma^2}{\Omega^2},\, -\frac{4\delta \omega_0^2}{\Omega^2},\, \frac{1}{2}\Omega t \right)\,,
\\
x_2(t) &= e^{-\frac{1}{2}\gamma t} \mathrm{S} 
\left( \frac{4\omega_0^2-\gamma^2}{\Omega^2},\, -\frac{4\delta \omega_0^2}{\Omega^2},\, \frac{1}{2}\Omega t \right)  \,,
\label{eqn:Mathieu_function}
\end{align}
where 
$\mathrm{C}(a, b, z)$ and $\mathrm{S}(a, b, z)$ are the even and the odd
Mathieu functions, respectively, as defined in 
Mathematica~\cite{Mathematica} (Mathematica notations are $\mathrm{MathieuC}[a, b, z]$ and $\mathrm{MathieuS}[a, b, z]$).
These functions are normalized as $\mathrm{C}(a, 0, z) = \cos(\sqrt{a}z)$, $\mathrm{S}(a,0,z) = \sin(\sqrt{a}z)$ for $b = 0$.

\begin{figure}
\includegraphics[width=3.0 in]{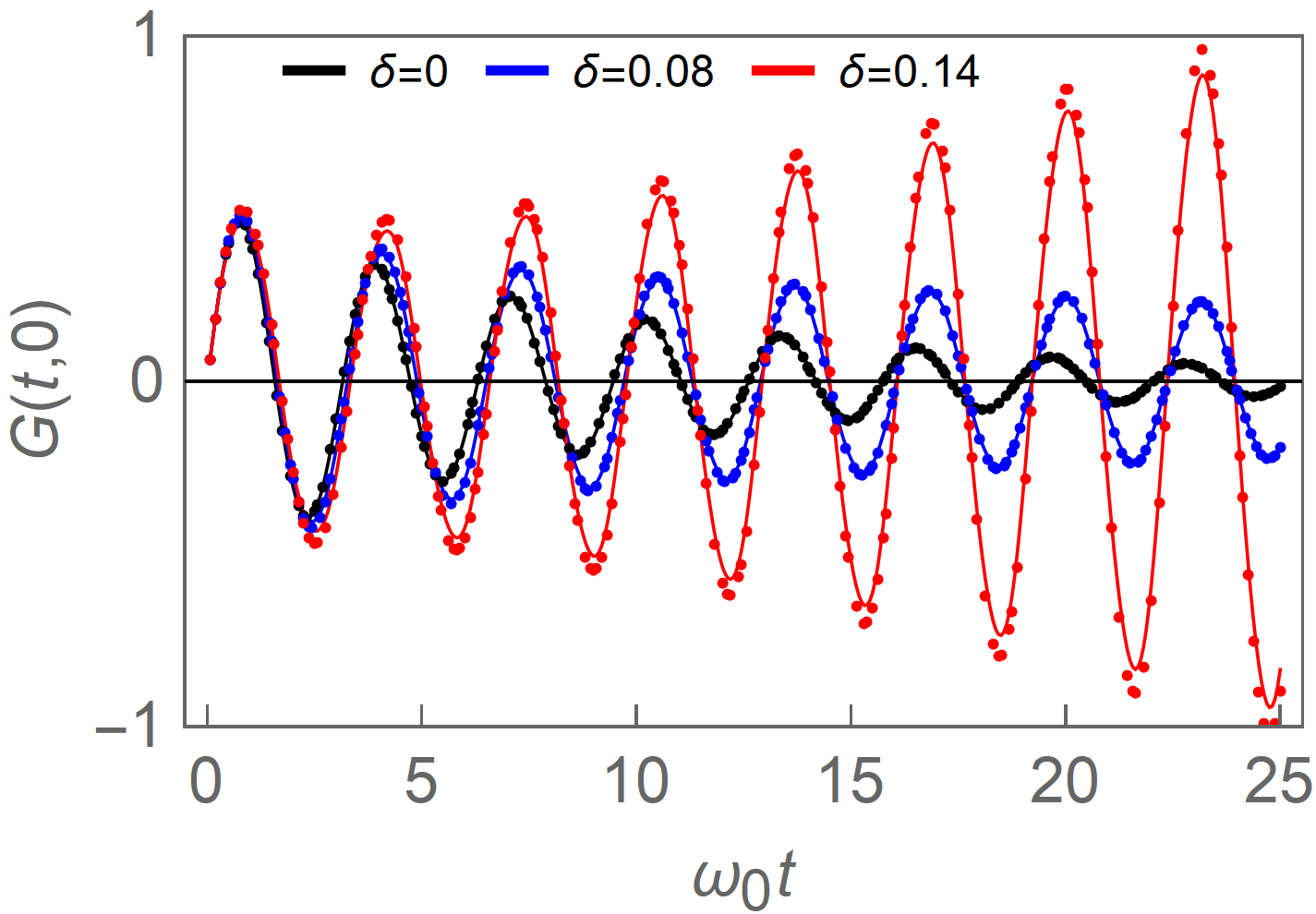} 
\caption{Behavior of the Green's function $G(t, 0)$ of Eq.~\eqref{eqn:OPO} at several values of the modulation parameter $\delta$ at resonance $\Omega = 2\omega_0$.
The solid lines show the exact Green's function [Eq.~\eqref{eqn:G}], the dots are the approximations constructed from Eq.~\eqref{eqn:approx_solution}.
The damping rate is $\gamma = 0.1 \omega_0$ corresponding to $\delta_c \approx 0.1$. 	
}
\label{fig:G}
\end{figure}

From Eqs.~\eqref{eqn:G}--\eqref{eqn:Mathieu_function}, we conclude that
\begin{align}
G(t, 0) = \frac{x_2(t)}{\dot{x}_2(0)}\,,
\label{eqn:G_t0}
\end{align}
so that an instability, i.e., the exponential growth of $G(t, 0)$
is possible if $\mathrm{S}(a, b, z)$ is increasing faster than $e^{\frac{1}{2} \gamma t}$.
This occurs if $|\delta|$ exceeds some threshold value
$\delta_c = \delta_c(\gamma, \Delta)$.
Conversely, the response function $G(t, 0)$
exhibits exponential decay with time
if $|\delta| < \delta_c$.
Examples of such stable and unstable behaviors are shown in Fig.~\ref{fig:G}.

Approximate analytical calculation of $\delta_c$ 
can be done for low damping $\gamma \ll \omega_0$ and
small detuning $\Delta \ll \omega_0$.
Thus, one can show that
\begin{equation}
\delta_c \simeq \frac{\sqrt{\Delta^2 + \gamma^2}}{\omega_0}\,,
\label{eqn:delta_c}
\end{equation}
in agreement with Fig.~\ref{fig:G}.

The derivation goes as follows.
Let us seek a solution of Eq.~\eqref{eqn:OPO} in the form
\begin{align}
x(t) = A(t) e^{i\omega_0 t} + B(t) e^{-i\omega_0 t} \,
\label{eqn:separate_amplitude}
\end{align}
where $A(t)$, $B(t)$ are slowly varying.
Matching the coefficients for the rapidly oscillating phase factors of Eq.~\eqref{eqn:OPO},
we obtain
\begin{equation}
\begin{pmatrix}
2 i \partial_t + i\gamma 		&& \omega_0 \delta e^{i\Delta t} \\
\omega_0 \delta  e^{-i\Delta t}	&& -2 i \partial_t - i \gamma
\end{pmatrix}
\begin{pmatrix}
A \\
B
\end{pmatrix} 
=
\begin{pmatrix}
0 \\
0
\end{pmatrix} 
.
\label{eqn:amplitude}
\end{equation}
To eliminate the factors $e^{\pm i\Delta t}$, we change variables to
\begin{align}
A(t) = e^{\frac{i}{2}\Delta t} a(t)\,,
\quad
B(t) = e^{-\frac{i}{2}\Delta t} b(t)\,,
\label{eqn:change_variable}
\end{align}
arriving at 
\begin{equation}
\begin{pmatrix}
2 i \partial_t + i \gamma - \Delta	&& \omega_0 \delta \\
\omega_0 \delta						&& -2 i \partial_t - i \gamma -\Delta
\end{pmatrix}
\begin{pmatrix}
a \\
b
\end{pmatrix} 
=
\begin{pmatrix}
0 \\
0
\end{pmatrix} 
.
\label{eqn:amplitude2}
\end{equation}
The general solution of this equation is
\begin{equation}
\begin{pmatrix}
a \\
b
\end{pmatrix} 
=
e^{\lambda_1 t} 
\begin{pmatrix}
u_1 \\
v_1
\end{pmatrix} 
+
e^{\lambda_2 t} 
\begin{pmatrix}
u_2 \\
v_2
\end{pmatrix} 
\,,
\label{eqn:solution1}
\end{equation}
where 
\begin{align}
\lambda_{1, 2} = -\frac{\gamma}{2} \pm \frac{i}{2} \sqrt{\Delta^2 - (\omega_0 \delta)^2}
\end{align}
are the growth rates and
\begin{align}
\begin{pmatrix}
u_{j} \\
v_{j}
\end{pmatrix} 
= 
\begin{pmatrix}
\omega_0 \delta
\\[0.5 em]
\Delta - i \gamma - i \lambda_j
\end{pmatrix},
\quad j = 1, 2 
\label{eqn:solution2}
\end{align}
are the corresponding eigenvectors.
Two linearly independent solutions for $x(t)$ could be chosen as
\begin{align}
x_j = e^{\lambda_j t} 
\left( u_j  e^{i \frac{\Omega}{2} t} + v_j  e^{-i \frac{\Omega}{2} t}  \right),
\quad j = 1, 2\,,
\label{eqn:approx_solution}
\end{align}
and the Green's function can be constructed per Eq.~\eqref{eqn:G}.
Assuming $\real{\lambda_2} \geq \real{\lambda_1}$,
the instability threshold is determined by the condition
\begin{align}
\real{\lambda_2} = 0\,,
\end{align}
which leads to Eq.~\eqref{eqn:delta_c}.
A comparison with the exact Green's function
confirms the validity of this approximation for small damping and detuning,
such as $\gamma = 0.1 \omega_0$ and $\Delta = 0$ in Fig.~\ref{fig:G}.

Consider now the response of the PO to a periodic driving source
\begin{equation}
F(t) = f_\omega e^{-i \omega t} + \text{c.c.} \,.
\label{eqn:F}
\end{equation}
In the stable case, $|\delta| < \delta_c$, where the effect of the initial
conditions disappears at long times, 
$x_\omega$ is the linear response to $F$:
\begin{equation}
x_\omega(t) = f_\omega e^{-i \omega t} \int\limits_0^{\infty}  G(t, t - \tau)
e^{i \omega \tau} d \tau
+ \text{c.c.} \,.
\label{eqn:PO_periodic}
\end{equation}
Without the pump, $G(t, t - \tau)$ is a function of $\tau$ only, and so
the integral on the right-hand side reduces to a constant.
Under pumping, $G(t, t - \tau)$ is periodic in $t$ with the period $2\pi / \Omega$.
Accordingly, the Fourier spectrum of $x_\omega(t)$ becomes a frequency comb
of Floquet harmonics $\omega + m \Omega$ where $m$ is an arbitrary integer.
For small $\gamma$ and $\Delta$, however, it is easier to compute $x_\omega(t)$
not from Eq.~\eqref{eqn:PO_periodic} but from Eq.~\eqref{eqn:amplitude2} supplemented with the appropriate right-hand side.
The result is
\begin{align}
x_\omega(t) =& \frac{f_\omega e^{-i\omega t}}{(\omega - \nu_1)(\omega - \nu_2)}
\left(
\frac{\Omega - 2\omega - i\gamma + \Delta}{4 \omega_0}
- \frac{\delta}{4} e^{i\Omega t}
\right)
\notag\\
&+ \text{c.c.} 
\,.
\label{eqn:pumped_kernel}
\end{align}
Apparently, this approximation accounts only for the dominant Floquet harmonics: $\omega - \Omega$, $\omega$,  and $\omega + \Omega$. In the complex plane of $\omega$, function $x_\omega$ (or the linear response kernel) has two poles, at $\nu_1$ and $\nu_2$,
given by
\begin{align}
\nu_{1, 2} = \frac{\Omega}{2} + i \lambda_{1, 2}
= \omega_0 + \frac{\Delta}{2} - \frac{i\gamma}{2}
\mp \frac{1}{2}\sqrt{\Delta^2 - (\omega_0 \delta)^2}\,.
\end{align}
The pole $\nu_2$ has a larger imaginary part, which grows with $|\delta|$. 
At $|\delta| = \delta_c$, $\nu_2$ crosses over into the upper half-plane
of $\omega$, signaling the instability.  

Instead of a periodic driving source, we can consider a stochastic one, namely, a
Langevin  source defined by the two-point correlator
\begin{align}
\big\langle F(t) F(0) \big\rangle = \frac{\gamma T }{\pi} \delta(t)\,.
\end{align}
In this case the total driving force $F(t)$ contains a continuum of Fourier harmonics,
and the response $x(t)$ is obtained by integrating expressions like
Eq.~\eqref{eqn:pumped_kernel} over $\omega$.
The stochastic drive leads to random fluctuations of $x$, causing
a nonzero average power injection by the parametric pump into the system:
\begin{align}
P =  2 \omega_0^2 \delta \left\langle\cos \Omega t\, x \partial_t x \right\rangle\,.
\label{eqn:P_pump}
\end{align}
A straightforward calculation based on Eqs.~\eqref{eqn:pumped_kernel} and \eqref{eqn:P_pump} yields
\begin{align}
P = \frac{({\omega_0 \delta})^2 \gamma T}{\gamma^2 + \Delta^2 - (\omega_0 \delta)^2 }\,.
\label{eqn:P_classical}
\end{align}

Finally, we generalize this result to the quantum case.
We do not show a formal derivation, but sketch a more physical approach. The essential point is that the power injection $P$ by the pump is enabled by the nonzero quantum/thermal fluctuations of $x$.
The starting point is a Hamiltonian of the total system (PO plus environment)
in the form
\begin{align}
H(t) &= H_{\mathrm{PO}} - F(t)  x + H_{\mathrm{env}}\,,
\label{eqn:H}\\
H_{\mathrm{PO}} &= \frac{p^2}{2} 
 + \frac{1}{2} \left( 1 + 2\delta \cos \Omega t \right) \omega_0^2 x^2\,.
\label{eqn:H_CL}
\end{align}
The purpose of $H_{\mathrm{env}}$ 
is to describe the Langevin noise source and the dissipation effects.
A standard device to achieve this is
to use the Caldeira-Leggett model
of a bath of harmonic oscillators.
We surmise that the final answer for $P$
is similar to the classical Eq.~\eqref{eqn:P_classical} except
the thermal energy $T$ is replaced by $\hbar\omega_0 \left(n_b + \frac12\right)$ where $n_b = 1 / (e^{\hbar \omega_0/T} - 1)$ is the boson occupation number.
Since the injected power increases the energy of the PO, the pump must be generating
quanta of the PO motion.
Dividing $P$ by the energy $2\hbar \omega_0$ of two such quanta, we obtain the pair generation rate
\begin{align}
R = \frac{P}{2\hbar\omega_0} = 
\frac{\gamma}{\gamma^2 + \Delta^2 - (\omega_0 \delta)^2}\,
\frac{({\omega_0 \delta})^2}{2}
\left(n_b + \frac12\right) \,.
\label{eqn:R_PO}
\end{align}
Note that formally, in the quantum mechanical language, this nonperturbative result in  the parametric pump $\delta$ can be obtained by computing the linear response  to $\delta$ (bubble diagram)~\cite{Golez.2020}, with the boson propagators replace by the renormalized ones by $\delta$ [see \equa{eqn:pumped_kernel}].	
As shown by \equa{eqn:R_PO}, at weak pumping, $R$ scales quadratically with $\delta$, consistent with Fermi's Golden Rule.
The precise match is obtained in the limit $\gamma \to 0$, $\omega_0 \delta / \gamma \to 0$
where $R$ is nonzero only at the resonance, $R \propto \delta(\Delta)$.
As $\delta$ increases,
deviations from Fermi's Golden Rule appear.
As $\delta$ approaches the critical value $\delta_c$ [Eq.~\eqref{eqn:delta_c}],
the pair generation rate diverges.

Lastly, expressing $\delta$ in terms of the coupling constant $g^{(2)}$ in Sec.~\ref{sec:main_results} and integrating over the acoustic modes in the vicinity of $\omega_0$ 
we obtain Eq.~\eqref{eqn:R}.

\section{Derivation for the graphene ribbon (type-A device)}
\label{sec:derivation}

\subsection{Plasmon modes on a generic nano structure}
\label{appendix:nano_structure}
Unlike the uniform two-dimensional (2D) electron gas, a generic conducting nano structure~\cite{Abajo.2007, Wang:2015vc, Manzoni} 	breaks the translational symmetry in at least one direction, and the plasmon modes can not be classified by 2D momenta. In this section, we define necessary profile functions describing the graphene nano island (e.g. a ribbon) such that the quantization could be described semi-analytically. Meanings of the dimensionless profile functions are summarized in Table~\ref{tbl:profiles}.

The linear optical conductivity is assumed local and in the Drude form: $\mathbf{j}(\mathbf{r})=\hat{\sigma}(\mathbf{r})\mathbf{E}(\mathbf{r})=\frac{i}{\pi (\omega+i\gamma)} D(\mathbf{r}) \mathbf{E}(\mathbf{r})$ where $D(\mathbf{r})$ is the local Drude weight. This holds for $\omega \gg v_F q$ where $q$ is the characteristic momentum corresponding to the plasmon modes and size of the nano structure. In the case of a graphene ribbon with $a$ being its half width, this momentum is roughly $q \approx \pi/(2a)$.  
We define the dimensionless conductivity profile function $\tilde{\sigma}$ by $\hat{\sigma}(\mathbf{r})=\hat{\sigma} \tilde{\sigma}(\mathbf{r})$ (or $D(\mathbf{r})=D \tilde{\sigma}(\mathbf{r})$) where $D$ is the maximum local Drude weight. Assume a certain plasmon mode has the charge density profile $\rho(\mathbf{r})=\rho \tilde{\rho}(\mathbf{r})$, its electric field profile is 
\begin{align}
	\mathbf{E}(\mathbf{r}) = - \nabla \hat{V} \rho(\mathbf{r}) = \rho \tilde{\mathbf{E}}(\mathbf{r}) \,, \quad \tilde{\mathbf{E}}(\mathbf{r}) \equiv - \nabla \hat{V}  \tilde{\rho}(\mathbf{r})
	\label{eqn:efield_profile}
\end{align}
where the Coulomb kernel $\hat{V}(\mathbf{r},\mathbf{r^{\prime}})$ is defined as an operator such that $\hat{V} f(\mathbf{r})= \int d\mathbf{r}^{\prime 2} \frac{1} {|\mathbf{r}-\mathbf{r^{\prime}}|} f(\mathbf{r^{\prime}})$.
Note that $\tilde{\sigma}$, $\tilde{\rho}$ and $\tilde{\mathbf{E}}$ are all dimensionless profile functions whose values are order one. 

From the charge continuity equation (CCE)
\begin{align}
	\partial_t \rho(\mathbf{r}) + \nabla \cdot \mathbf{j} = 0 \,
	\label{eqn:EOM_island}
\end{align}
with only linear conductivity, the mode $\rho(\mathbf{r})=\rho \tilde{\rho}(\mathbf{r})$ being an eigenmode with frequency $\omega_1$ implies
\begin{align}
	-i \omega_1 \tilde{\rho}(\mathbf{r}) - \sigma(\omega_1) \nabla \tilde{\sigma}(\mathbf{r}) \nabla \hat{V} \tilde{\rho}(\mathbf{r}) = 0 \,.
	\label{eqn:EOM_island}
\end{align}
Define the linear operator $\hat{L}=\nabla \tilde{\sigma}(\mathbf{r}) \nabla \hat{V}$, the above equation simplifies to
\begin{align}
	\hat{L} \tilde{\rho}(\mathbf{r}) = \frac{-i\omega_1}{\sigma(\omega_1)} \tilde{\rho}(\mathbf{r})  \,,
	\label{eqn:L}
\end{align}
i.e., the shape function $\tilde{\rho}(\mathbf{r})$ is an eigen vector of $\hat{L}$ with eigen value $-i\omega_1 / \sigma(\omega_1)$. We define inner product to be $\langle f|g\rangle \equiv \int f^{\ast}(r)\hat{V}(r,r^\prime)g(r^\prime)d^2r d^2r\prime$ such that $\hat{L}$ is hermitian. Although the plasmon frequency is complex with a small imaginary part $\mathrm{Im}[\omega_1]=\gamma/2$, the eigen value $-i\omega_1 / \sigma(\omega_1)$ is real since $\hat{L}$ is a hermitian operator.

If an external uniform electric field $\mathbf{E}_{\text{ext}}=E_{\text{ext}} \mathbf{e}$ with frequency $\omega$ is applied, the CCE becomes
\begin{align}
	(-i \omega -  \sigma(\omega) \hat{L})\tilde{\rho}(\mathbf{r})  \rho  + \sigma(\omega) \nabla \tilde{\sigma}(\mathbf{r}) \mathbf{e} E_{\text{ext}}  = 0 \,.
\end{align}
Taking inner product with $\tilde{\rho}(\mathbf{r})$, we arrive at
\begin{align}
	\left( -i \omega -  \sigma(\omega) \frac{-i\omega_1}{\sigma(\omega_1)}
	\right)
	|\tilde{\rho}|^2  \rho  + \sigma(\omega) \alpha E_{\text{ext}}  = 0 \,
\end{align}
where 
\begin{align}
	|\tilde{\rho}|^2 &=\langle \tilde{\rho}|\tilde{\rho}\rangle = \int \tilde{\rho}(r) \hat{V}(r,r^\prime) \tilde{\rho}(r^\prime)d^2r d^2r\prime \notag\\
	& =  \frac{\sigma(\omega_1)}{i\omega_1} \int \tilde{\sigma}(r) |\tilde{\mathbf{E}}(\mathbf{r})|^2 d^2r = \frac{\sigma(\omega_1)}{i\omega_1} S 
	\label{eqn:a_alpha}
\end{align}
and the dipole factor is
\begin{align}
	\alpha &=\langle \tilde{\rho}|  \mathbf{e} \cdot \nabla \tilde{\sigma}   \rangle =  \mathbf{e} \cdot \int \tilde{\rho}(r)\hat{V}(r,r^\prime) \nabla^\prime \tilde{\sigma} (r^\prime)  d^2r d^2 r\prime \notag \\
	&= \mathbf{e} \int  \tilde{\sigma}(r) \tilde{\mathbf{E}}(r)  d^2r  \,.
\end{align}
Two new shape quantities are introduced: $S$ has the interpretation of the effective area~\cite{Manzoni} of this mode, $\alpha$ is the effective dipole moment of the mode along $\mathbf{e}$. Therefore, the charge density amplitude induced by the external driving field is \equa{eqn:EOM_source}.
In this response function, there is a simple pole at resonance $\omega=\mathrm{Re}[\omega_1]$, leading to divergence. However, this pole is broadened by the plasmon damping rate $\gamma$, and the enhancement factor at resonance will be
\begin{align}
	\rho=(Q_1/i) (\alpha/S) E_{\text{ext}}
	\label{eqn:rho_resonance}
\end{align} 
where $Q_1=\omega_1/\gamma$ is the quality factor of this plasmon mode.

\begin{table}
	\begin{ruledtabular}
		\begin{tabular}{ll}
			Symbol  & Physical quantity
			\\
			\hline
			$\tilde{\sigma}(\mathbf{r})$ & conductivity
			\\[6pt]
			$\tilde{g}(\mathbf{r})$ & 2nd-order conductivity $\sigma^{(2)}$
			\\[6pt]
			$\tilde{\rho}(\mathbf{r})$ & charge density of a mode
			\\[6pt]
			$\tilde{\mathbf{E}}(\mathbf{r})$ & electric field of a mode
		\end{tabular}
	\end{ruledtabular}
	\caption{Physical meanings of the dimensionless profile functions whose typical local values are order unity.}
	\label{tbl:profiles}
\end{table}

\subsection{Quantization}
\label{sec:Quantization}
For each plasmon mode as a harmonic oscillator, we choose charge density $\rho$ as the generalized coordinate and write it in terms of plasmon creation and annihilation operators as
\begin{align}
	\rho = \rho_u (a+a^\dagger) \,.
	\label{eqn:rho_quantization}
\end{align}
Equivalently (Eq.~\eqref{eqn:efield_profile}) the electric field is
\begin{align}
	\mathbf{E}(\mathbf{r}) = \rho_u (a+a^\dagger) \tilde{\mathbf{E}}(\mathbf{r}) \,.
	\label{eqn:E_quantization}
\end{align}
Working in the gauge $A=(0,\mathbf{A})$ for the vector potential, from the relation $\mathbf{E}=-\partial_t \mathbf{A}/c$, we can write the vector potential as 
\begin{align}
	\mathbf{A}(\mathbf{r})/c = -i\frac{\rho_u}{\omega_1} (a-a^\dagger) \tilde{\mathbf{E}}(\mathbf{r}) \,
	\label{eqn:A_quantization}
\end{align}
where the operators are time dependent ones from the interaction picture $a(t)=a e^{-i\omega_1 t}$.

Since the potential energy of a harmonic oscillator is half of its total energy, the quantum unit of density $\rho_u$ can be determined by   
\begin{align}
	\langle V_{\text{Coulomb}} \rangle= \frac{1}{2} \rho_u^2 |\tilde{\rho}|^2=\frac{1}{4} \hbar \omega_1
	\label{eqn:vCoulomb}
\end{align}
which yields
\begin{align}
	\rho_u=\sqrt{\hbar\omega_1 / (2|\tilde{\rho}|^2) } \,.
	\label{eqn:rho_u}
\end{align}

We note that due to nonzero damping, the modes here should be understood as quasi-normal modes~\cite{Franke.2019,Ren.2021, Sauvan.2022}. Nevertheless, due to the nice properties of Drude response for plasmons, we were able to cast the formalism into a Hermitian one with a simpler approach (\equa{eqn:L})  such that the basis functions are automatically orthonormal. The quantization procedure assumes no dissipation, which is then added phenomenologically and should be understood as coming implicitly from a bath.

\subsection{Three mode coupling due to second-order nonlinearity}

In general, the current generated in response to electric field contains a second-order term
\begin{align}
	j^{(2)}_i(\mathbf{r}, t) = \hat{\sigma}^{(2)}_{ilm} [E_{l}(\mathbf{r}, t) E_{m}(\mathbf{r}, t)] \,,
\end{align}
where $\hat{\sigma}^{(2)}_{ilm}$ is a tensor-valued operator nonlocal in space and time.
In a uniform system, this operator is diagonal in the frequency-momentum space, so that
\begin{align}
	j^{(2)}_i(-\mathbf{q}_3, -\omega_3) = \sigma^{(2)}_{ilm}
	E_{l}(\mathbf{q}_1, \omega_1) E_{m}(\mathbf{q}_2, \omega_2) \,.
\end{align}
where $\omega_3 = -(\omega_1 + \omega_2)$, $\mathbf{q}_3 = -(\mathbf{q}_1 + \mathbf{q}_2)$, and
\begin{equation}
	\sigma^{(2)}_{ilm} =  \sigma^{(2)}_{ilm}(\mathbf{q}_1, \omega_1; \mathbf{q}_2, \omega_2)\,.
	\label{eqn:sigma2}
\end{equation}
By convention, the nonlinear second-order conductivity $\sigma^{(2)}_{ilm}$ is symmetrized, i.e.,
invariant under the interchange $(1 \leftrightarrow 2, l \leftrightarrow m)$.
It is convenient to define another tensor ${\Pi}_{ilm}$, which 
describes the current response to the vector potential $\mathbf{A}$.
In the temporal gauge, $\mathbf{E}(\mathbf{q}, \omega) = (i \omega / c) \mathbf{A}(\mathbf{q}, \omega)$,
it is given by
\begin{align}
	\hat{\Pi}_{ilm}(\omega_3,\mathbf{q}_3;\omega_1, \mathbf{q}_1,\omega_2,\mathbf{q}_2)
	\equiv -\omega_1 \omega_2 {\sigma}^{(2)}_{ilm} \,,
\end{align}
so that
\begin{align}
	j^{(2)}_i(-\mathbf{q}_3, -\omega_3) = \frac{1}{c^2}\,
	\hat{\Pi}_{ilm} A_l(\mathbf{q}_1, \omega_1) A_m(\mathbf{q}_2, \omega_2)\,.
\end{align}

Due to inversion symmetry of graphene,
$\sigma^{(2)}_{ilm}$ vanishes at ${q}_1 = {q}_2 = 0$.
It grows linearly with ${q}_1$, ${q}_2$
when these momenta are small.
The same properties are inherited by the tensor  ${\Pi}_{ilm}$. 
In particular, in the kinetic regime of graphene~\cite{Mikhailov2011, Tokman2016, Cheng2016, Sun2018a} and neglecting dissipation,
${\Pi}_{ilm}$ can be written as
\begin{align}
	\hat{\Pi}_{ilm} =& -\frac{ D^{(2)}}{\omega_1 \omega_2 \omega_3}  
	\bigg[\big( \omega_2^2 q_1 + \omega_1^2 q_2
	+ 2\omega_1\omega_2 q_3 \big)_i \delta_{lm} 
	+  \mathop{\text{perm}}
	\bigg]  ,
	\label{eqn:Pi}
\end{align}
[When comparing to other expressions in the literature, one should remember the constraint $\sum_k \omega_k = \sum_k \mathbf{q}_k = 0$.]
The notation ``$\mathop{\text{perm}}$'' stands for two additional terms obtained from the first one by the permutations
\begin{align}
	(3,i;1,l;2,m) &\rightarrow (1,l;3,i;2,m)\,,
	\notag\\
	(3,i;1,l;2,m) &\rightarrow (2,m;1,l;3,i)\,.
\end{align}
Accordingly, $\hat{\Pi}_{ilm}$ is symmetric under arbitrary permutations of the triads $\{\mathbf{q}_k, \omega_k, \mu_k\}$, where $k = 1$, $2$, $3$ and $\mu_k = i$, $l$, $m$.
This symmetry can be understood as a consequence of energy conservation.

The tensor ${\Pi}_{ilm}$ defines the operator $\hat{\Pi}_{ilm}$ nonlocal in space and time.
The three-mode coupling Hamiltonian can be constructed as 
\begin{align}
	H_c= -\frac{1}{3 c^3}\int d^2r \hat{\Pi}_{ilm} A_i A_l A_m\,.
	\label{eqn:Hc}
\end{align}
This Hamiltonian obeys the requirement
$\partial H_c /\partial {\mathbf{A}} =-\mathbf{j}^{(2)}/c$. Assume any three plasmon modes $1$, $2$, and $3$, Eq.~\eqref{eqn:Hc} together with Eq.~\eqref{eqn:A_quantization} lead to their coupling Hamiltonian
\begin{align}
	H_c 
	=&-2i \frac{\rho_{1u} \rho_{2u} \rho_{3u}}{\omega_0 \omega_2 \omega_3}  \int d^2r \hat{\Pi}_{ilm} \tilde{E}_{1i} \tilde{E}_{2l} \tilde{E}_{3m} \notag\\
	& \cdot (a_1-a_1^\dagger)(a_2-a_2^\dagger)(a_3-a_3^\dagger) \,.
	\label{eqn:Hc_Q}
\end{align}
In $\hat{\Pi}$, the momenta  should be replaced by spatial gradients acting on the corresponding field profiles, and the frequency arguments  should match those of the creation/annihilation operators (e.g., for terms like $a_1 a_2^\dagger a_3^\dagger$, it should be  $\hat{\Pi}(-\omega_0,\omega_2,\omega_3)$). 

If the nanostructure has inversion symmetry, the plasmon modes can have even or odd parity. In order for the interaction Eq.~\eqref{eqn:Hc_Q} not to vanish, the product of the three fields must have even parity. In the case of subharmonic decay, the mode $0$ should couple to light whose electric field is nearly uniform, and is thus an odd mode.
The other two modes should be nearly identical whose product is thus even, leading the product of the three modes to be odd.
Therefore, the desired subharmonic decay does not happen in inversion symmetric nano structures.
We investigate the approach to break the inversion symmetry in the next section.

\subsection{The graphene ribbon}
\label{appendix:ribbon}
For the graphene ribbon shown in Fig.\ref{fig:pn}, there is mirror symmetry of $x \rightarrow -x$ while the static transverse electric field breaks the mirror symmetry of $y \rightarrow -y$. The gapped mode $1$ and the acoustic mode $0$ with momentum $q$ along $x$ have the electric potential
\begin{align}
	\varphi_{1} = \varphi_1(y) \,, \quad 
	\varphi_{0} = \varphi_0(y) e^{i q x} \,.
	\label{eqn:acoustic_mode_potential}
\end{align}
Fully extracting the dependence on length scales, the effective areas can be written as 
\begin{align}
	S_{0} &= aL S_0^\prime = aL \int_{-1}^{1} dy \tilde{\sigma}(y) |\mathbf{\tilde{E}}_0|^2 \,, \notag \\
	S_{1} &= aL S_1^\prime = aL \int_{-1}^{1} dy \tilde{\sigma}(y) |\mathbf{\tilde{E}}_1|^2 \,,
	\label{eqn:S}
\end{align}
where $S_0^\prime$ and $S_1^\prime$ are dimensionless factors of order one. The dipole factor for the gapped mode is
\begin{align}
	\alpha =aL\alpha^\prime = aL \int_{-1}^{1} dy \tilde{\sigma}(y) \hat{\mathbf{e}}\cdot \mathbf{\tilde{\mathbf{E}}}_1  \,.
\end{align}
where $\alpha^\prime$ is order one.

In section~\ref{sec:Quantization},  we have assumed that the density shape functions $\tilde{\rho}(\mathbf{r})$, as transformation coefficients for the `generalized coordinates' $\rho$, are all real functions. Due to the translational invariance along $x$, it is convenient to define the `complex' modes $\rho_q$ with well defined momenta. In the conventional quantization rule $\rho_q = \rho_u (a_q+a_{-q}^\dagger)$,  the creation operator $a^\dagger_q$ generates plasmon with momentum $q$, i.e., its action adds $\hbar q$ to the total momenta. This set of creation and annihilation operators are related to those of Eq.~\eqref{eqn:rho_quantization} by a canonical transformation.

\emph{The interaction strength}---
If the chemical potential on this ribbon varies slowly enough: $1/\lambda_\mu \ll \omega/v_F$ where $\lambda_\mu$ is the characteristic spatial scale of chemical potential variation, one can make the local approximation, using Eq.~\eqref{eqn:Pi} as the local second order nonlinear coupling strength. 
The second order optical weight can be written as $D^{(2)}=D_k^{(2)} \tilde{g}(y)$ where $D_k^{(2)}$ is a typical value of $D^{(2)}$ on the strip and $\tilde{g}(y)$ is an order one dimensionless shape function. Given fixed chemical potential, the value of $D_k^{(2)}$ is temperature dependent~\cite{Sun2018a}. In the high frequency kinetic regime, $D_k^{(2)}$ goes to a constant value
\begin{align}
	D_0^{(2)}= -\frac{1}{8 \pi}\frac{e^3 v^2}{\hbar^2} \mathrm{sign}\, \mu \,
\end{align}
at $T \ll \mu$ and vanishes as $T \gg \mu$.
Since $\mu \gg T$ at the edge of the strip, we define $D_k^{(2)}$ to be its zero temperature value $D_0^{(2)}$. 

With Eqs.~\eqref{eqn:Pi}, \eqref{eqn:S} and \eqref{eqn:rho_u},
the coupling Hamiltonian Eq.~\eqref{eqn:Hc_Q} applied to modes $0$, $q$ and $-q$ becomes
\begin{align}
	H_c 
	=g^{(2)}_{q} a_1 a_{q}^\dagger a_{-q}^\dagger +c.c. \,
	\label{eqn:Hc_Q2}
\end{align}
where we have kept only the near resonant terms. Therefore, we have derived the interaction term in Eq.~\eqref{eqn:hamiltonian} where the interaction strength $g$ is given by Eq.~\eqref{eqn:g} with $\xi_g = \xi_1^{1/2} \eta^{\prime}/(S_1^{\prime 1/2} S_0^\prime)$ and the dimensionless integral $\eta^\prime$ defined as
\begin{align}
	\eta^\prime= q_x^{\prime 2} \int_{-1}^{1} dy \, 
	\partial_y^2 \varphi_1 \,
	\varphi_0^2 \,
	\tilde{g}(y)  \,.
	\label{eqn:eta_prime}
\end{align}
In the above expression for $\eta^\prime$, we have dropped the terms involving $\partial_y \varphi_0$. This approximation is good if $\partial_x \varphi_0 \gg \partial_y \varphi_0$ which is true for the subharmonic plasmons in Fig.~\ref{fig:pn}(b). The dimensionless momentum $q_x^{\prime}=q_x a$ is order one.

In Fig.~\ref{fig:shape_factor}, we plot the resulting shape factors for the strip as the linear doping profile varies. There we have assumed the temperature and doping dependence of the Drude weight and  the second order optical weight are $D(\mu, T)=2\frac{e^2}{\hbar^2} T \ln \left[2 \cosh \left(\frac{\mu}{2T} \right)\right]$ and $D_k^{(2)}(\mu, T)=D_0^{(2)} \tanh\left( \frac{\mu}{2T}\right)$, so that their dimensionless profile functions are $\tilde{\sigma}(y)=\frac{2T}{\mu(y)} \ln \left[2 \cosh \left(\frac{\mu(y)}{2T} \right)\right]$ and $\tilde{g}(y)=\tanh\left( \frac{\mu(y)}{2T}\right)$.

\emph{The occupation number}---Under the AC electric field of the incident light, the gapped mode $\omega_1$ is driven to a coherent state $|\beta(t)\rangle$ where $\beta(t)=\beta e^{-i\omega_1 t}$. From Eq.~\eqref{eqn:rho_quantization}, the average value of $\rho$ is  
\begin{align}
	\langle \rho \rangle = \rho_u \langle \beta(t) |(a+a^\dagger)|\beta(t)\rangle = \rho_u (\beta e^{-i\omega_1 t} + c.c.) \,
	\label{eqn:rho_average}
\end{align}
which combined with Eq.~\eqref{eqn:rho_resonance} yields
\begin{align}
	\beta = (Q/i) (\alpha/S) E_{ext}/\rho_u \,.
	\label{eqn:beta2}
\end{align}
The above equation and Eq.~\eqref{eqn:rho_u}, \eqref{eqn:gapped_dispersion} yields Eq.~\eqref{eqn:beta}.

\subsection{The decay rate}
\label{sec:decay_rate}
\begin{figure}
	\includegraphics[width=0.6 \linewidth]{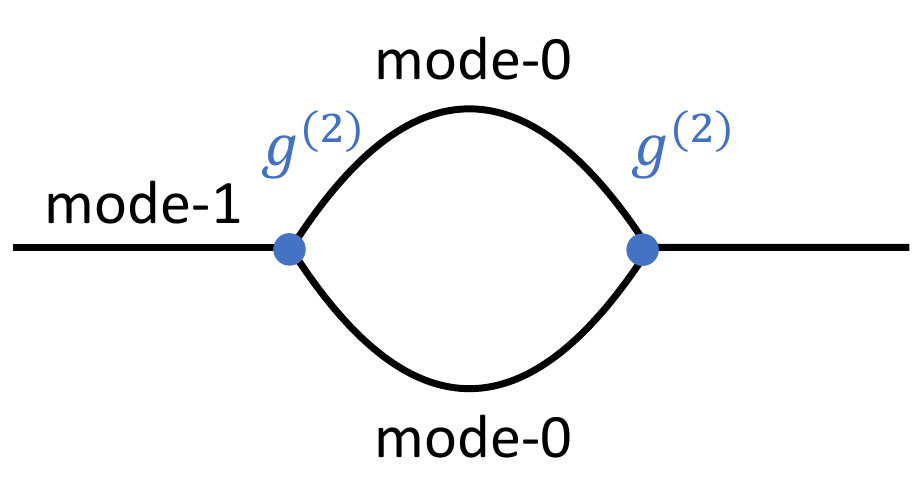} 
	\caption{Feynman diagram for the self energy of  mode-1 plasmon due to its parametric coupling to two mode-0 plasmons. Solid lines are the plasmon propagators.}
	\label{fig:parametric_decay}
\end{figure}
In this subsection, we derive \equa{eqn:decay_rate}, the decay rate $\Gamma$ of the mode-1 plasmon into two mode-0 plasmons through the interaction Hamiltonian \equa{eqn:hamiltonian}. This could be done by calculating the self energy (\fig{fig:parametric_decay}) of the mode-1 plasmon. It is a bubble diagram formed by two mode-0 propagators:
\begin{align}
\Sigma(i\omega)=&\left|\frac{g^{(2)}}{\hbar}\right|^2  \sum_{q, i\Omega_n} \frac{ \omega(q)}{(i\Omega_n)^2-\omega(q)^2}
\frac{\omega(q)}{(i\omega+i\Omega_n)^2-\omega(q)^2}
\notag\\
=& \left|\frac{g^{(2)}}{\hbar}\right|^2 \sum_q 
\frac{\left[2n(\omega(q))+1\right]  4\omega(q)}{(i\omega)^2-4\omega(q)^2}
\,.
\end{align}
where $n(x)= 1/(e^{\hbar x/ T} - 1)$ is the boson occupation number. 
Its imaginary part at $i\omega=\omega_1$ would simply give the Fermi's golden rule:
\begin{align}
	\mathrm{Im}[\Sigma(\omega_1)]
	=&\left|\frac{g^{(2)}}{\hbar}\right|^2 \sum_q \left[2n(\omega(q))+1\right] 
	\pi\delta(\omega_1-2\omega(q))	
	\notag\\
	=& \left|\frac{g^{(2)}}{\hbar}\right|^2 \frac{L}{v_0} \left(n(\omega_1/2)+\frac{1}{2}\right) 
	\,
\end{align}
which is just \equa{eqn:decay_rate}.

\section{Classical parametric amplification via second-order nonlinearity}
\label{sec:second_order}
\begin{figure}
	\includegraphics[width=\linewidth]{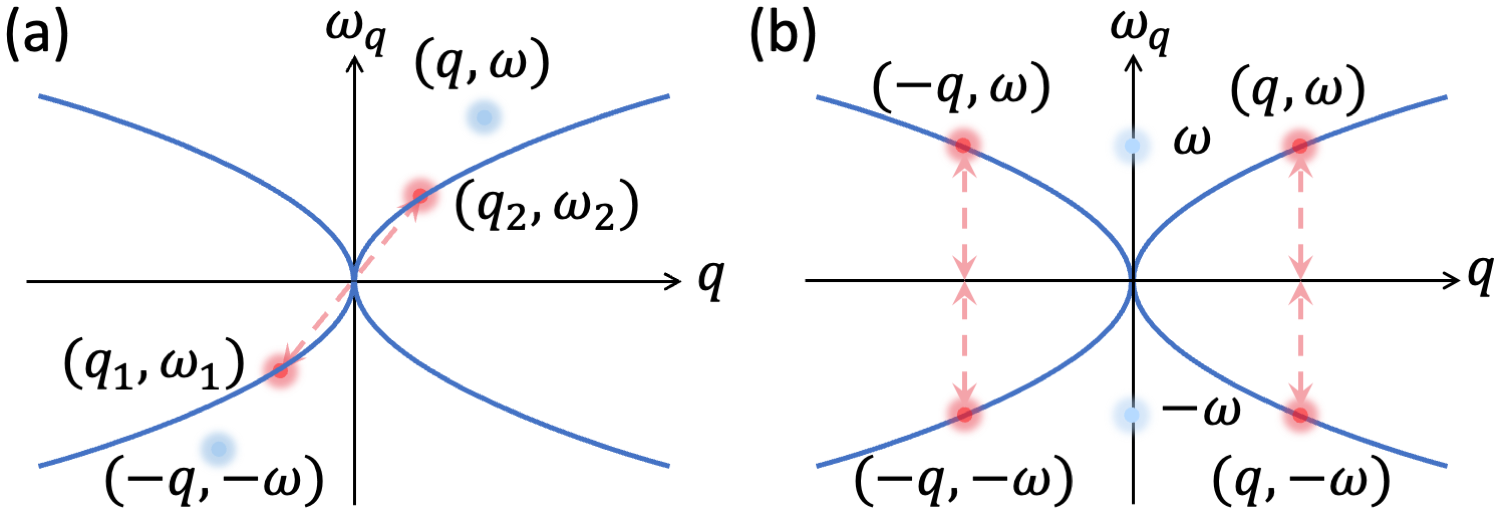} 
	\caption{(a) Schematic illustration of 2D plasmons parametrically driven by the pump through second order nonlinearity. The blue curve is the plasmon frequency momentum dispersion. The red dots are the two plasmon modes coupled by the pump field at $(q,\omega)$ and its complex conjugate  at $(-q,-\omega)$. (b) Degenerate FWM of plasmons induced by third order nonlinearity. The dots at zero momentum represent the uniform pump field with frequency $\omega$ and it complex conjugate. The dots at the same frequency but nonzero momentum represent the plasmon modes coupled by this field through third order nonlinearity.}
	\label{fig:second_order_instability}
\end{figure}

\subsection{Infinite plane}
\emph{General}---For the subharmonic generation to happen in inversion symmetric systems, there needs to be a strong pump field with frequency $\omega$ and appreciable wave vector $\mathbf{q}$ whose electric potential is
\begin{align}
\Phi=\phi e^{i(\mathbf{q} \mathbf{r} -\omega t )} + \phi^{\ast} e^{-i(\mathbf{q} \mathbf{r} -\omega t )} \,.
\label{eqn:driving_potential}
\end{align}

The equation of motion (EOM) of the plasmons is conveniently described by the second order nonlinear conductivity
\begin{align}
j_\mu(\omega,\mathbf{q}) &=\int d\omega^{\prime} d\mathbf{q}^{\prime} \sigma_{\mu\alpha\beta}^{(2)} (\omega-\omega^{\prime},\mathbf{q}-\mathbf{q}^{\prime},\omega^{\prime},\mathbf{q}^{\prime})
\notag \\
& E_\alpha(\omega-\omega^{\prime},\mathbf{q}-\mathbf{q}^{\prime}) E_\beta(\omega^{\prime},\mathbf{q}^{\prime})  \,.
\label{eqn:nonlinear_conductivity}
\end{align}
Since $\sigma^{(2)}$ is a real function in space-time domain, its frequency momentum representation has the general property
\begin{align}
\sigma_{\mu\alpha\beta}^{(2)} (-\omega_1,-\mathbf{q}_1,-\omega_2,-\mathbf{q}_2) = \sigma_{\mu\alpha\beta}^{{(2)}\ast} (\omega_1,\mathbf{q}_1,\omega_2,\mathbf{q}_2)
\,.
\label{eqn:sigma2_general_property}
\end{align}
And if the system has inversion symmetry, then
\begin{align}
\sigma_{\mu\alpha\beta}^{(2)} (\omega_1,-\mathbf{q}_1,\omega_2,-\mathbf{q}_2) = - \sigma_{\mu\alpha\beta}^{(2)} (\omega_1,\mathbf{q}_1,\omega_2,\mathbf{q}_2)
\,
\label{eqn:inversion_symmetry}
\end{align}
which means
$
\sigma_{\mu\alpha\beta}^{(2)} (\omega_1,0,\omega_2,0) = 0
\,.
\label{eqn:inversion_symmetry2}
$
However, it can be nonzero at finite momentum.
By convention, $\sigma^{(2)}$ is defined to be symmetric: 
$\sigma_{\mu\alpha\beta}^{(2)} (\omega_1,\mathbf{q}_1,\omega_2,\mathbf{q}_2) = \sigma_{\mu\beta\alpha}^{(2)} (\omega_2,\mathbf{q}_2,\omega_1,\mathbf{q}_1)$.

Due to $\sigma^{(2)}$ and the driving field, the equation of motion for plasmons is modified. As illustrated in Fig.~\ref{fig:second_order_instability}(a), any two modes $(\mathbf{q}_1, \mathbf{q}_2)$ with $\mathbf{q}_1 + \mathbf{q} = \mathbf{q}_2$ are coupled by the driving field through $\sigma^{(2)}$. The charge continuity equation for the mode with momentum $\mathbf{q}_2$ is 
\begin{align}
\partial_t \rho_{\mathbf{q}_2} + \nabla \cdot \mathbf{j}_{\mathbf{q}_2} =0 \,
\label{eqn:charge_continuity}
\end{align}
and the current $\mathbf{j}_{\mathbf{q}_2}$ can be expanded to second order in electric field
\begin{align}
j_\mu(\omega_2,\mathbf{q}_2) = \sigma_{\mu\nu}(\omega_2,\mathbf{q}_2) E^{\nu}(\omega_2,\mathbf{q}_2) + \notag \\
\sigma_{\mu\alpha\beta}^{(2)} (\omega_1,\mathbf{q}_1,\omega,\mathbf{q}) E^{\alpha}(\omega_1,\mathbf{q}_1) E^{\beta}(\omega,\mathbf{q}) + \notag \\
\sigma_{\mu\alpha\beta}^{(2)} (\omega,\mathbf{q},\omega_1,\mathbf{q}_1)  E^{\alpha}(\omega,\mathbf{q}) E^{\beta}(\omega_1,\mathbf{q}_1)
\,.
\label{eqn:jk2}
\end{align}
The electric field $E(\omega_2,\mathbf{q}_2)$ is related to the charge density through the Coulomb kernel
\begin{align}
\mathbf{E}(\omega_2,\mathbf{q}_2) = -i \mathbf{q}_2 V_{q_2} \rho(\omega_2,\mathbf{q}_2) \,.
\label{eqn:efield_rho}
\end{align}
Similar set of equations apply to the pairing mode $\rho_{\mathbf{q}_1}$. Separating the densities into products of amplitude and phase as 
\begin{align}
\rho_{\mathbf{q}_1} = A_1(t) e^{i(\mathbf{q}_1 \mathbf{r} -\omega_1 t )} ,\quad
\rho_{\mathbf{q}_2} = A_2(t) e^{i(\mathbf{q}_2 \mathbf{r} -\omega_2 t )}
\,,
\label{eqn:rho_separate}
\end{align}
and plugging them into the continuity equations (Eq.~\eqref{eqn:charge_continuity} et al), we get the equations for time evolution equation of the amplitudes:
\begin{equation}
\begin{pmatrix}
\partial_t + \gamma && \kappa_2 e^{i\Delta t} \\
\kappa_1 e^{-i\Delta t}  && \partial_t +\gamma
\end{pmatrix}
\begin{pmatrix}
A_1 \\
A_2
\end{pmatrix} 
=
0 \,.
\label{eqn:rho_amplitude}
\end{equation}
The parameters $\kappa_1$, $\kappa_2$ are 
\begin{align}
\kappa_1 &=iq_{2 \mu}  \sigma_{\mu\alpha\beta}^{(2)} (\omega_1,\mathbf{q}_1,\omega,\mathbf{q}) 
(-iq_{1 \alpha} V_{q_1})(-iq_{\beta} \phi)  \notag \\
& + iq_{2 \mu}  \sigma_{\mu\alpha\beta}^{(2)} (\omega,\mathbf{q},\omega_1,\mathbf{q}_1) (-iq_{\alpha} \phi)
(-iq_{1 \beta} V_{q_1}) \notag \\
&= -i 2 \phi V_{q_1} \sigma_{\mu\alpha\beta}^{(2)} (\omega_1,\mathbf{q}_1,\omega,\mathbf{q}) q_{2 \mu} q_{1 \alpha} q_{\beta} 
\,, \notag \\
\kappa_2 
&= i 2 \phi^{\ast} V_{q_2} \sigma_{\mu\alpha\beta}^{(2)} (\omega_2,\mathbf{q}_2,-\omega,-\mathbf{q}) q_{1 \mu} q_{2 \alpha}   q_{\beta}   \,.
\label{eqn:kappa} 
\end{align}
Simmilar to Eq.~\eqref{eqn:amplitude}, the solution to Eq.~\eqref{eqn:rho_amplitude} has an exponential factor $e^{\lambda t}$. For there to be nonzero solutions, we require the determinant to be zero, and obtain
\begin{align}
\lambda = - \gamma \pm \sqrt{\kappa_1 \kappa_2 - \Delta^2/2 } \,
\label{eqn:determinant}
\end{align}
where $\Delta=\omega-(\omega_2-\omega_1)$. Therefore, the criterion for there to be an exponentially growing solution is 
\begin{align}
\mathrm{Re}\left[\sqrt{\kappa_1 \kappa_2  - \Delta^2/2 }\right] > \gamma \,.
\label{eqn:criterion}
\end{align}
Let's temporarily assume that there is no damping $\gamma=0$. In the ideal case, $\mathbf{q}_1=-\mathbf{q}_2$, and $\Delta=0$, we have $\kappa_2=\kappa^{\ast}_1$ from the property Eq.~\eqref{eqn:sigma2_general_property}. Therefore, Eq.~\eqref{eqn:criterion} is guaranteed to be satisfied in any dissipation-less system.

\emph{Applied to Graphene}---In the kinetic regime of graphene, expanding to linear order in $q$, the second order nonlinear conductivity reads~\cite{Wang2016,Sun2018a}
\begin{widetext}
\begin{align}
&\sigma^{(2)}_{ilm}(\omega_1, \mathbf{q}_1,\omega_2,\mathbf{q}_2)
= \frac{D_k^{(2)}}{\omega_1 \omega_2 \omega_3} \Bigg[ \left(q_3 -\frac{\omega_1}{2\omega_2}q_2 -\frac{\omega_2}{2\omega_1}q_1 \right)_i \delta_{lm}
+ \left(2 q_2 - 2 q_1 - \frac{\omega_2}{\omega_1} q_1 +\frac{3\omega_1}{\omega_2}q_2 \right)_m \delta_{il}
\Bigg]
+	
\left(\begin{smallmatrix}
1 & \leftrightarrow& 2\\ l &\leftrightarrow& m
\end{smallmatrix} \right)\, 
\label{eqn:sigma2_graphene_boltzmann_2}
\end{align}
\end{widetext}
where $(\omega_3,\mathbf{q}_3)=(\omega_1+\omega_2,\mathbf{q}_1+\mathbf{q}_2)$
and the second order optical weight is
$D_k^{(2)} = D_0^{(2)} \tanh \left(\mu/(2 T)\right)$~\cite{Sun2018a}
whose zero temperature limit is $D^{(2)}_0$. Applied to Eq.~\eqref{eqn:kappa} in the ideal case $\mathbf{q}_1=-\mathbf{q}_2$, and $\Delta=0$, we obtain the growth rate
\begin{align}
\sqrt{\kappa_1 \kappa_2} = \frac{3}{2} \frac{e^3 v^2_F}{\hbar^2} \frac{q^3}{\omega^3} |\phi|  \,.
\label{eqn:growth_optimal_case} 
\end{align}
Taking into account the plasmon dispersion $\omega_1=\sqrt{2D q_1}$, we arrive at the dimensionless relative growth factor
\begin{align}
Q_g^{-1} = \frac{3}{16 \alpha_g}  \zeta \,.
\label{eqn:growth_q_factor}
\end{align}

\subsection{Graphene ribbon}
Assume the $\omega_1$ mode is excited by an external source tuned exactly at $\omega_1$, the density is $\rho(t)=\rho_0 e^{-i\omega_1 t}+c.c.$. Through second order nonlinearity, the strong field of mode $\omega_1$ is going to modify the CCE of the $\omega_0$ mode. Assuming the charge density of the subharmonic mode is $\rho_1 =A(t) e^{-i\omega_0 t} + c.c. $, plugging it into the CCE of this mode and taking inner product with $\tilde{\rho}_1$,
one obtains equation for the amplitude:
\begin{align}
\partial_t A + \kappa A^\ast =0 \,
\label{eqn:amplitudes_coupled}
\end{align}
where $|\tilde{\rho}_1|^2$ is defined in Eq.~\eqref{eqn:a_alpha} and 
$
|\kappa|=|\beta g/\hbar|
$.
Eq.~\eqref{eqn:amplitudes_coupled} leads to exponentially growing solution with the relative growth rate 
\begin{align}
Q^{-1}_g = \frac{|\kappa|}{\omega_0} = \zeta \frac{Q }{8 \alpha_g} \frac{ \eta^\prime \alpha^\prime }{\xi^2_0 S_{1}^\prime S_0^\prime} \,
\label{eqn:growth_rate_island}
\end{align} 
which agrees with Eq.~\eqref{eqn:Qg}, the quantum mechanical result.

\section{Classical parametric amplification via third-order nonlinearity}
\label{sec:third_order_apd}
Assume a strong uniform electric field $\mathbf{E}=\mathbf{E}_0 e^{-i\omega t} + c.c.$ which, due to momentum mismatch, cannot excite plasmons through linear response. However, through third order nonlinear effect, this uniform $\mathbf{E}$ tends to make the plasmons grow in amplitude. The CCE of the plasmons with momentum $q$ is
\begin{align}
& \partial_t \rho_{\mathbf{q}} + \nabla \cdot \mathbf{j}_{\mathbf{q}} =0 \,,
\label{eqn:EOM}
\end{align}
and the currents are 
\begin{align}
j(\mathbf{q},\omega)_{i} =& \sigma_{il}(\omega) E(\mathbf{q},\omega)_l +
\notag\\
&3\sigma^{(3)}_{ilmn}(-\omega,\omega,\omega) E(\mathbf{q},-\omega)_l E_{0m} E_{0n} \,, \notag\\
j(\mathbf{q},-\omega)_{i} =& \sigma_{il}(-\omega) E(\mathbf{q},-\omega)_l +
\notag\\
&3\sigma^{(3)}_{ilmn}(\omega,-\omega,-\omega) E(\mathbf{q},\omega)_l E^{\ast}_{0m} E^{\ast}_{0n}
\label{eqn:EOM}
\end{align}
where
$E(\mathbf{q})_l = -i q_l v_q \rho_{\mathbf{q}}$. Thus the CCEs of the plasmon mode $q$ at frequencies $\omega$ and $-\omega$ are coupled by the pump field $E_0$, as shown in Fig.~\ref{fig:second_order_instability}(b).
Separating amplitude and phase of the charge density as
\begin{align}
\rho_{\mathbf{q}}(t) = A(t) e^{-i\omega t} + B(t) e^{i\omega t} \,
\label{eqn:EOM}
\end{align}
we arrive at the equation for the amplitudes
\begin{equation}
\begin{pmatrix}
2\partial_t  && \lambda \\
\lambda^{\prime}  && 2\partial_t
\end{pmatrix}
\begin{pmatrix}
A \\
B
\end{pmatrix} 
=
0 \,
\label{eqn:amplitude_third_order}
\end{equation}
where 
\begin{align}
\lambda &= 3 \sigma^{(3)}_{ilmn}(-\omega,\omega,\omega) q_i q_l v_q E_{0m} E_{0n} \,, \notag \\
\lambda^{\prime} &=  3 \sigma^{(3)}_{ilmn}(\omega,-\omega,-\omega) q_i q_l v_q E^{\ast}_{0m} E^{\ast}_{0n}  \,.
\label{eqn:EOM}
\end{align}
Due to the fact that $\sigma^{(3)}_{ilmn}(\omega,-\omega,-\omega)=\sigma^{(3)\ast}_{ilmn}(-\omega,\omega,\omega)$, we have $\lambda^{\prime}=\lambda^{\ast}$.
Therefore, Eq.~\eqref{eqn:amplitude_third_order} has the solution
\begin{align}
A(t) &= A_+ e^{\kappa t} + A_- e^{-\kappa t} ,
\notag\\
B(t) &= B_+ e^{\kappa t} + B_- e^{-\kappa t} ,
\label{eqn:solution}
\end{align}
where $\kappa = |\lambda|$ is the growth/decay rate. Taking the `Drude' form in \equa{eqn:sigma_3} for $\sigma^{(3)}$ of graphene in the kinetic regime, plugging in the square root dispersion of plasmons $\omega=\sqrt{2 D q }$, the dimensionless relative growth rate $Q_g^{-1}=\kappa/\omega$ simplifies to 
\begin{align}
Q_g^{-1} =  \frac{3}{8}  \frac{e^2 E^2_0 /\omega^2 }{ ( \hbar  k_F )^2} = \frac{3}{8}   \xi^2
\end{align}
for the plasmons propagating parallel to the direction of the pump field.
The dimensionless small number $\xi = \frac{e E_0 /\omega}{ \hbar  k_F } = \delta p /p_F $ is simple to memorize: $\delta p$ is just the change in electron momentum caused by the electric field during one half cycle of the oscillations, $\delta t \sim \pi / \omega$~\cite{Sun2018b}. Note that due to third order nonlinearity, the plasmon frequency itself is also renormalized by this strong uniform field~\cite{Sun2018b}, and the frequency $\omega$ throughout this section should be considered as the renormalized one. 

Although a result of third order nonlinearity, this phenomenon is different from modulational instability which is ubiquitous in nonlinear optics and fluid mechanics~\cite{Falkovich2011}, e.g. in surface gravity waves. In modulational instability, the strong pump is a finite momentum `wave train' on resonance, and the exponentially growing waves have momentums different from the pump by a small amount $\delta k$. If the criterion of instability is satisfied, the growth rate scales linearly with $\delta k$. For a negative Kerr nonlinearity as for graphene plasmons~\cite{Sun2018b}, the criterion requires $\partial^2_k \omega > 0$, not satisfied by the square root dispersion. However, this criterion assumes a non dispersive $\sigma^{(3)}$ which is not true in graphene. Therefore, whether modulational instability happens in graphene needs further investigation and is a weaker effect anyway. Indeed, in a recent work on nonlinear plasmons~\cite{Eliasson2016}, it was found that second order nonlinearity could lead to growth of side bands in the presence of a wave train, similar to modulational instability.

\section{Near-field probe of DFG}
\label{appendix:DFG_near_field}
In this section, we discuss a scanning near field experiment that could measure plasmonic DFG by classical interference. The setup is similar to the left part of Fig.~\ref{fig:SNOM_q}.
Upon pumping of the $\omega_1$ mode of the device Fig.~\ref{fig:pn}(a), if the $\omega_0$ mode with momentum $q$ is launched by a classical source combined with an `antenna', e.g., the left edge of the ribbon, the counter propagating mode with momentum $-q$ will be generated by DFG from the pump and the $q$ mode. The $q$ mode can be described by a coherent state $|a_0\rangle$ which satisfies $a_q |a_0\rangle=a_0 |a_0\rangle$. The Hamiltonian~\eqref{eqn:hamiltonian} leads to equation of motion for the $-q$ mode:
\begin{align}
(\partial_t + \gamma) \langle a_{-q} \rangle = -i\omega_0 \langle a_{-q} \rangle  -i g^{(2)} \beta a_0^\ast e^{-i\omega_0 t}/\hbar
\label{eqn:EOM_a_q}
\end{align}
where the phenomenological damping rate $\gamma$ has been added. After the driving term has been turned on for a time long enough, the steady state solution  is
\begin{align}
\langle a_{-q} \rangle  =  i \frac{g^{(2)} \beta}{\hbar \gamma} a_0^\ast e^{-i\omega_0 t} = i \frac{\kappa }{ \gamma} a_0^\ast e^{-i\omega_0 t} \,.
\label{eqn:a_q_amplitude}
\end{align}
Therefore, if the `reflection coefficient' $\kappa / \gamma$ is at the order of one, the two waves would interfere to form fringes with period $\lambda_0/2$ where $\lambda_0=2\pi/q$ is the wavelength of the subharmonic plasmon. These fringes could be picked up by the near field scanning probe.

\bibliography{./nonlinear_plasmon}

\end{document}